\documentclass[aps,preprint]{revtex4}
\usepackage{amsmath}
\usepackage{amssymb}
\usepackage{mathrsfs}
\usepackage{xcolor}
\usepackage{graphicx}
\usepackage{subfigure}
\usepackage{enumerate}
\usepackage{float}

\begin{document}
	\title{Joule-Thomson expansion of the lower-dimensional black hole in rainbow gravity}

	\author{Siyuan Hui$^{a,b}$}
	\email{huisiyuan@stu.scu.edu.cn}
	\author{Benrong Mu$^{a,b}$}
	\email{benrongmu@cdutcm.edu.cn}
	\author{Jun Tao$^{b}$}
	\email{taojun@scu.edu.cn}
	
	\affiliation{$^{a}$Physics Teaching and Research section, College of Medical Technology, Chengdu University of Traditional Chinese Medicine, Chengdu, 611137, PR China\\
		$^{b}$Center for Theoretical Physics, College of Physics, Sichuan University, Chengdu, 610064, PR China}

	\begin{abstract}
	
	In this paper, we extend Joule-Thomson expansion to the low-dimensional regime in rainbow gravity by considering the rainbow rotating BTZ metric in the (2+1)-dimensional spacetime. After the metric of the black hole is obtained, we get the Joule-Thomson expansion of the black hole, including the Joule-Thomson coefficient, inversion curves, and isenthalpic curves. We find that a rainbow rotating BTZ black hole does not have $P-V$ critical behavior. The effects of rainbow gravity are to slow down the trend of the increase of the Joule-Thomson coefficient and make its zero point larger. Moreover, the rainbow gravity slows down the inverse temperature of the black hole, meaning that a rainbow rotating BTZ black hole tends to change its heating or cooling action at a lower temperature, which can be attributed to the topology of the black hole.

	\end{abstract}
	\maketitle
	\tableofcontents
	
	\section{Introduction}
	
	In 1974, Stephen Hawking found that the Schwarzschild black hole emits radiation at a temperature just like an ordinary blackbody. Named as the Hawking radiation\cite{Hawking:1975vcx}, it was a very remarkable prediction of the theory about thermodynamics of black holes. Since then, black hole thermodynamics has been an area of great interest, where various thermodynamic properties of black holes have been extensively investigated\cite{Bekenstein:1972tm, Bekenstein:1973ur, Bardeen:1973gs, Bekenstein:1974ax, Hawking:1974rv, Hawking:1975vcx}. In 1982, Hawking and Page first obtained the thermodynamics of AdS black holes, and they found that there is a phase transition between a Schwarzschild AdS black hole and a thermal AdS space\cite{Hawking:1982dh}. Later, Hawking radiation and phase transition in different black holes attracted more attention\cite{Jiang:2005ba, Setare:2006hq, Jiang:2007pn, Myung:2006sq, Jiang:2007wj}.
	
	After investigating the phase transition\cite{Hawking:1982dh}, Hawking and Page found that a black hole in AdS space has similar thermodynamic properties as a general thermodynamic system. This similarity is further enhanced in extended phase space\cite{Kubiznak:2012wp}, which has been supported by the  $P-V$ criticality studied in extended phase space\cite{Ma:2017pap, Hendi:2012um, Wei:2015ana, Xu:2015rfa, Mo:2014mba}. In extended phase space, the cosmological constant is considered as thermodynamic pressure
	\begin{equation}
		P=-\frac{\Lambda}{8 \pi}=\frac{(d-1)(d-2)}{16 \pi l^2},
	\end{equation}
	where $l$ is the AdS space radius and the mass of black hole $M$ is considered as the black hole enthalpy\cite{Kastor:2009wy, Kubiznak:2016qmn}. Various interesting studies have been carried out on thermodynamic aspects of black holes in extended phase space, such as the phase transition\cite{Altamirano:2013uqa, Wei:2014hba, Hendi:2017fxp}, critical phenomenon\cite{Wei:2012ui, Banerjee:2011cz, Niu:2011tb} and generalized Smarr relation\cite{Smarr:1972kt, Rasheed:1995zv, Sekiwa:2006qj, Kastor:2010gq, Kastor:2011qp}. \"Okc\"u and Ayd\i{}ner first studied the Joule-Thomson expansion of black holes\cite{Okcu:2016tgt}. It was soon extended to many other black holes, such as $d$-dimensional charged AdS black holes\cite{Mo:2018rgq}, RN-AdS black holes in $f(r)$ gravity\cite{Chabab:2018zix}, quintessence RN-AdS black holes\cite{Ghaffarnejad:2018exz}, regular (Bardeen)-AdS black holes\cite{Pu:2019bxf}, Bardeen-AdS black holes\cite{Li:2019jcd}, Kerr-AdS black holes\cite{Okcu:2017qgo} and others\cite{RizwanCL:2018cyb, MahdavianYekta:2019dwf, Cisterna:2018jqg, Lan:2018nnp, Cao:2021dcq, Meng:2020csd, Bi:2020vcg, Guo:2020zcr, Rostami:2019ivr, Haldar:2018cks, Mo:2018qkt, Nam:2019zyk, Ghaffarnejad:2018tpr, Kuang:2018goo, Zhao:2018kpz, Lan:2019kak, Guo:2019pzq, Ranjbari:2019ktp, Sadeghi:2020bon, Farsam:2020pfl, Liang:2021elg, Yin:2021akt, Belhaj:2018cdp, Liang:2021xny}.
	
	On the other hand, DSR theory makes the Planck length a new invariant scale and gives out the nonlinear Lorentz transformations in momentum spacetime\cite{Amelino-Camelia:2000cpa, Amelino-Camelia:2000stu, Magueijo:2001cr, Magueijo:2002am}.
	Specifically, the modified energy-momentum dispersion relation of particle energy $E$ and momentum $p$ can take the following form
	\begin{equation}
		E^2 f(E/E_p)^2-p^2 g(E/E_p)^2=m^2,
	\end{equation}
	where $E_{p}$ is the Planck energy.Amelino-Camelia\cite{Amelino-Camelia:1996bln,Amelino-Camelia:2008aez} proposed a popular choice of solution, which gives
	\begin{equation}
		\begin{aligned}
			f(x)=1,
			g(x)=\sqrt{1-\eta x^n}.
		\end{aligned}
	\end{equation}
	It is compatible with some results obtained in the loop quantum gravity method and reflects those obtained in $\kappa$-Minkowski spacetime and other noncommutative spacetimes. The phenomenological meaning of this "Amelino-Camelia (AC) dispersion relation" was also reviewed\cite{Amelino-Camelia:2008aez}. Subsequently, Magueijo and Smolin proposed that the spacetime background felt by a test particle depends on the test particle’s energy\cite{Magueijo:2002xx}. Thus, the energy of the test particle deforms the background geometry and eventually leads to modified dispersion relations. According to the modified dispersion relations under the generalized uncertainty principle, the second law of black hole thermodynamics is proved to be valid by modifying a relation between the mass and the temperature of the black hole\cite{Amelino-Camelia:2005zpp}. Many studies have been done to explore various aspects of black holes and cosmology\cite{Galan:2004st, Hackett:2005mb, Aloisio:2005qt, Amelino-Camelia:2013wha, Barrow:2013gia, Garattini:2011hy, Garattini:2011fs, Garattini:2014rwa, Mu:2019jjw, Gangopadhyay:2016rpl, Mu:2015qna, Gim:2015yxa, Ali:2014zea, Kim:2016qtp, Ling:2005bp, MahdavianYekta:2019dwf}.
	
	There have been some studies focused on thermodynamics of various black holes in rainbow gravity, but few of them study the effects of rainbow gravity have on Joule-Thomson expansion. Joule-Thomson expansion of charged AdS black holes in rainbow gravity has been studied and some novel physical phenomena have been found\cite{ MahdavianYekta:2019dwf}, but it is focused on the spacetime with dimension $d \geq 4$, whereas the case of $d < 4$ remains to be explored. Three-dimensional Einstein gravity is topological, which means that all geometries sourced by the same matter are locally identical, with difference lying in the topology. And the rotating BTZ black hole is one solution of the Einstein field equation in the (2+1)-dimensional spacetime, with its Joule-Thomson expansion being discussed\cite{Liang:2021xny}. However, to date, the effects of rainbow gravity to Joule-Thomson expansion of the rotating BTZ black hole has not been explored. In this study, we investigated the Joule-Thomson expansion of a rainbow rotating BTZ black hole, trying to clarify the role of topology in Joule-Thomson expansion under rainbow gravity.
	
	In this paper, we study the Joule-Thomson expansion of a rotating BTZ black hole in rainbow gravity. The remainder of our article is summarized as follow. In Section \ref{222}, the metric of the rotating BTZ black hole in rainbow gravity is obtained. In Section \ref{333}, the Joule-Thomson expansion of the black hole, including the Joule-Thomson coefficient, inversion curves, and isenthalpic curves, is discussed. In Section \ref{444}, the discussion and conclusion is given.Throughout the paper we take geometrized units $c = G = k_b = 1$.

	\section{Rotating BZT black hole in rainbow gravity}\label{222}
	
	There are two functions called rainbow function $f(x)$ and $g(x)$ with the following properties
	\begin{equation}
		\lim\limits_{x\to0}f(x)=1,\qquad \lim\limits_{x\to0}g(x)=1.
	\end{equation}
	
	Specifically, the modified energy-momentum dispersion relation of particle energy $E$ and momentum $p$ can take the following form:
	\begin{equation}
		E^2 f(E/E_p)^2-p^2 g(E/E_p)^2=m^2,
	\end{equation}
	where $E_{p}$ is the Planck energy,
	and we set
	\begin{equation}
		\frac{E}{E_{p}}\equiv x.
	\end{equation}
	
	The modified dispersion relation (MDR) might play an important role in astronomical and cosmological observations, such as the threshold anomalies of ultra-high energy cosmic rays and $TeV$ photons. In phenomenological physics, ground observations and astrophysical experiments have tested the predictions of MDR theory. One of the most popular choices for the functions $f(x)$ and $g(x)$ has been proposed by Amelino-Camelia\cite{Amelino-Camelia:1996bln, Amelino-Camelia:2008aez}, which gives
	\begin{equation}
		\begin{aligned}
			f(x)=1,
			g(x)=\sqrt{1-\eta x^2}.
		\end{aligned}
	\end{equation}
	
	The metric of the rotating BTZ black hole in rainbow gravity is\cite{Liang:2021xny, Amelino-Camelia:1996bln, Amelino-Camelia:2008aez}
	\begin{equation}
		ds^2 = -\frac{h(r)}{f(x)^2} dt^2 +\frac{1}{g(x)^2 h(r)} dr^2 + \frac{r^2}{g(x)^2} \left( d \phi^2 - \frac{J^2}{2 r^2} dt^2   \right),
	\end{equation}
	where
	\begin{equation}
		h(r) =-8 M+\frac{r^2}{l^2}+ \frac{J^2}{4 r^2}.
	\end{equation}
	Here, $J$ is the angular momentum. The solution of the equation $h(r) = 0$ represents the event horizon, located at $r = r_+$.
	
	In extended phase space, the function with the cosmological constant $\Lambda$ is considered as the pressure of the black hole\cite{Kastor:2009wy, Kubiznak:2016qmn}
	\begin{equation}\label{6}
		P = -\frac{\Lambda}{8 \pi}= \frac{(d-1) (d-2)}{8 \pi l^2} = \frac{1}{8 \pi l^2},
	\end{equation}
	and the corresponding thermodynamic quantities are as follow
	\begin{equation}\label{1}
		H\equiv M=\frac{r_{+}^2}{8 l^2}+\frac{J^2}{32 r_{+}^2},\quad V=\pi r_{+}^2,\quad \Omega=\frac{J}{16 r_{+}^2},\quad T=\left( \frac{r_{+}}{2 \pi l^2}-\frac{J^2}{8 \pi r_{+}^3} \right) g(x).
	\end{equation}
	In the extended phase space, the first law of thermodynamics is given by\cite{Wang:2006eb}
	\begin{equation}\label{2}
		dM = TdS + VdP + \Omega dJ,
	\end{equation}
	and the corresponding Smarr formula is
	\begin{equation}\label{3}
		2 PV= TS + \Omega J.
	\end{equation}

	\section{Joule-Thomson expansion}\label{333}

	In Joule-Thomson expansion, gas passes through a porous plug or a small valve from high-pressure section to low-pressure section, and the enthalpy remains constant during the process. Therefore, the Joule–Thomson expansion of a black hole is an isenthalpic process in the extended phase space. The Joule–Thomson coefficient $\mu$ describes the temperature variation with respect to pressure and characterizes the expansion process, which is given by
	\begin{equation}
		\mu = \left(\frac{\partial T}{\partial P} \right)_H.
	\end{equation}
	The Joule-Thomson coefficient is an important physical quantity to study the Joule-Thomson expansion. The expansion is characterized by the change of temperature relative to pressure, and its sign can be used to determine whether heating or cooling occurs. The change of pressure during expansion is negative, as the pressure is always decreasing. However, the change of temperature is uncertain. When the temperature change is negative, the coefficient $\mu$ will be positive, which means that the cooling occurs. Conversely, when the temperature change is positive, the coefficient $\mu$ will be negative, which means that the heating occurs.

	The entropy of a BTZ black hole in rainbow gravity has been given in \cite{Mu:2019jjw}. From equation(\ref{1}), (\ref{2}) and (\ref{3}), the heat capacity at constant pressure is
	\begin{equation}
		C_P= T \left(  \frac{\partial S}{\partial T}  \right)_{P,J} =\frac{\pi  r_{+} \left(\eta +r_{+}^2\right) \left(32 \pi  P r_{+}^4-J^2\right)}{2 \sqrt{\frac{r_{+}^2 \left(1-\eta  m^2\right)}{\eta +r_{+}^2}} \left(J^2 \left(2 \eta +3 r_{+}^2\right)+32 \pi  P r_{+}^4 \left(2 \eta +r_{+}^2\right)\right)},
	\end{equation}
	and we can obtain
	\begin{equation}
		\begin{aligned}
			\mu &= \left(\frac{\partial T}{\partial P} \right)_H= \frac{1}{C_P} \left[ T \left(  \frac{\partial V}{\partial T}  \right)_{P} - V \right] \\
			&= \frac{2 r_{+} \sqrt{\frac{r_{+}^2 \left(1-\eta  m^2\right)}{\eta +r_{+}^2}} \left(J^2 \left(4 \eta +5 r_{+}^2\right)-32 \pi  P r_{+}^6\right)}{\left(\eta +r_{+}^2\right) \left(J^2-32 \pi  P r_{+}^4\right)}.
		\end{aligned}
	\end{equation}
	
	\begin{figure}[htbp]
		\centering
		\subfigure[$\eta=0$]{
			\includegraphics[scale=0.54]{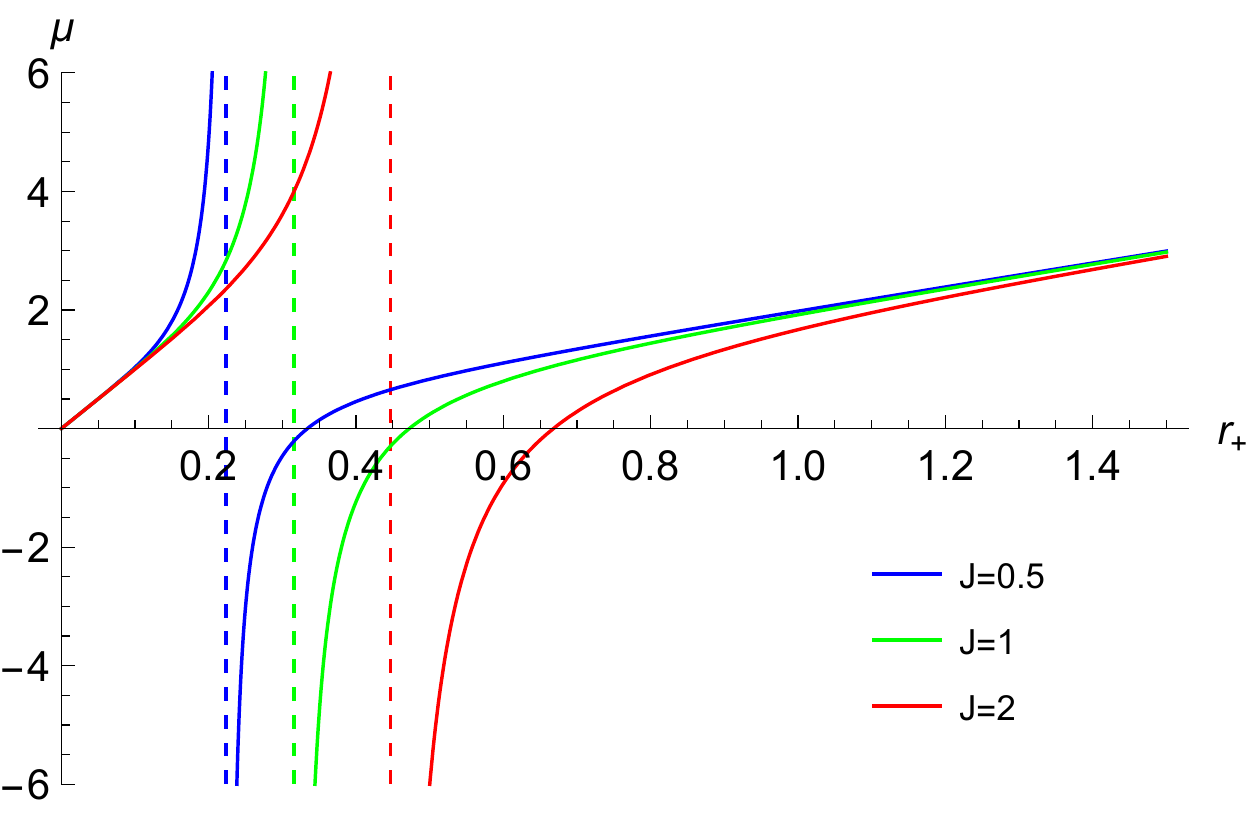}
			\includegraphics[scale=0.48]{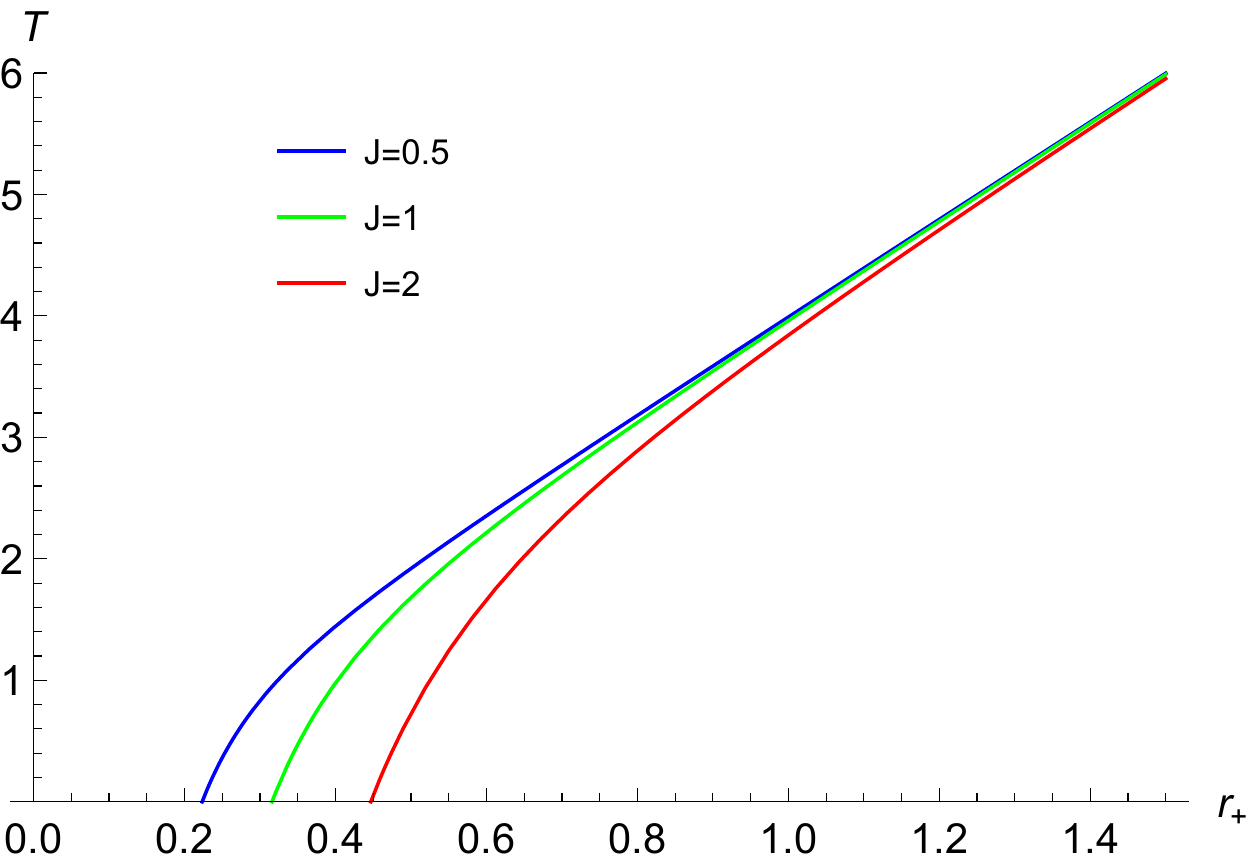}}
		\quad
		\subfigure[$\eta=0.1$]{
			\includegraphics[scale=0.54]{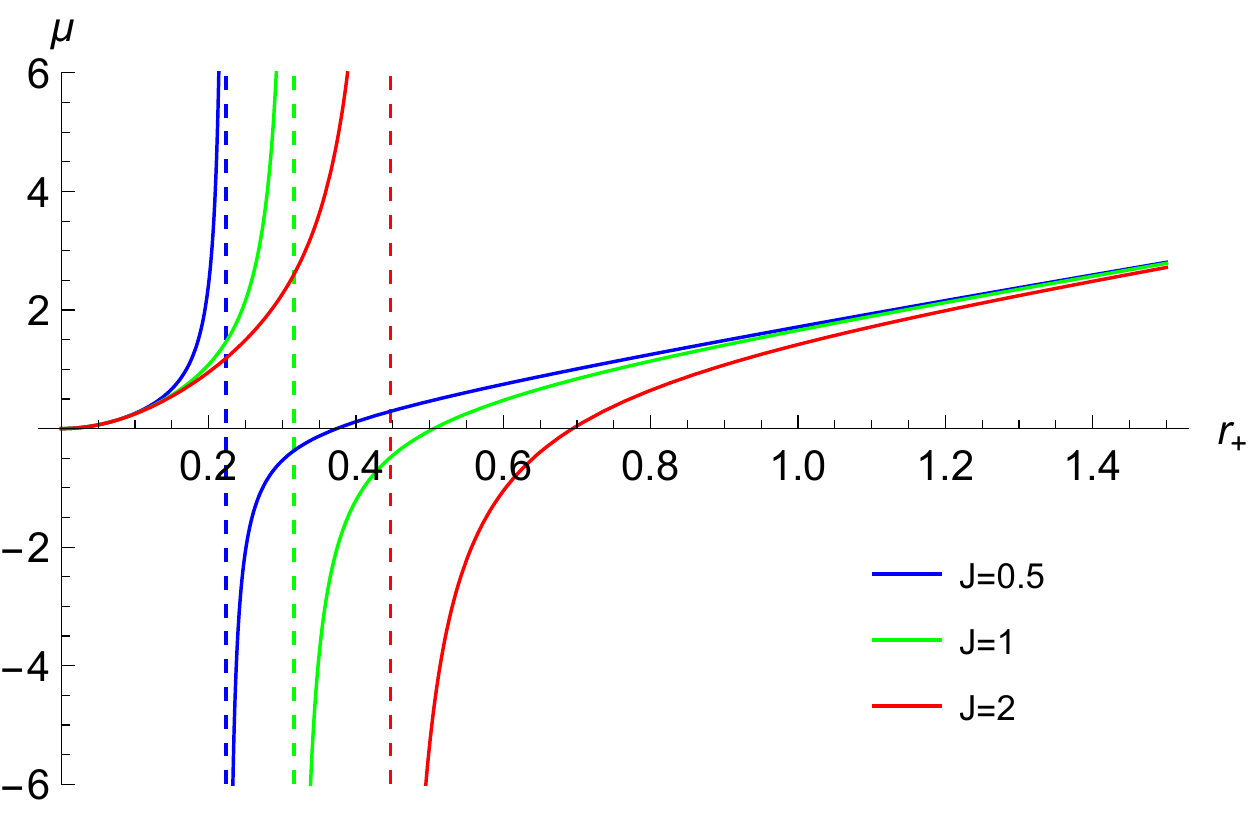}
			\includegraphics[scale=0.48]{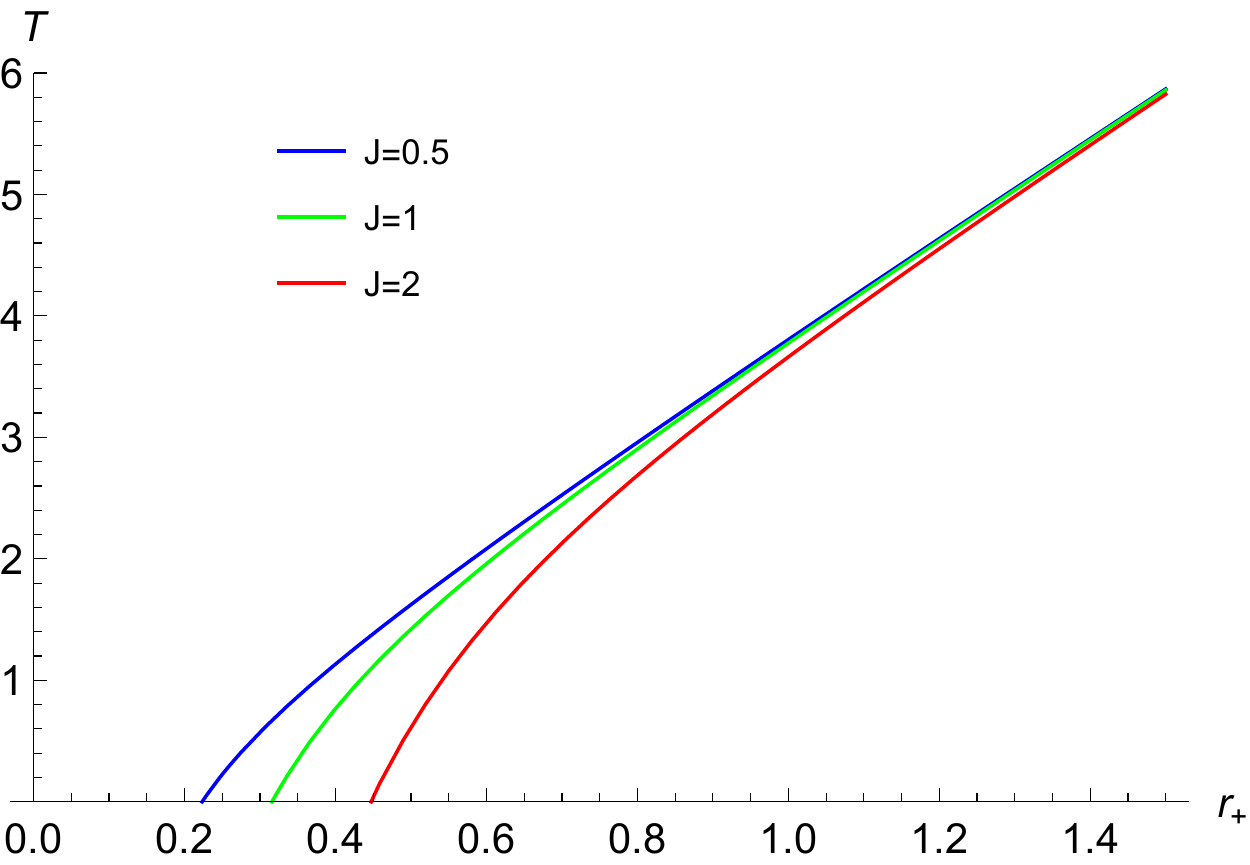}}
		\quad
		\subfigure[$\eta=0.2$]{
			\includegraphics[scale=0.54]{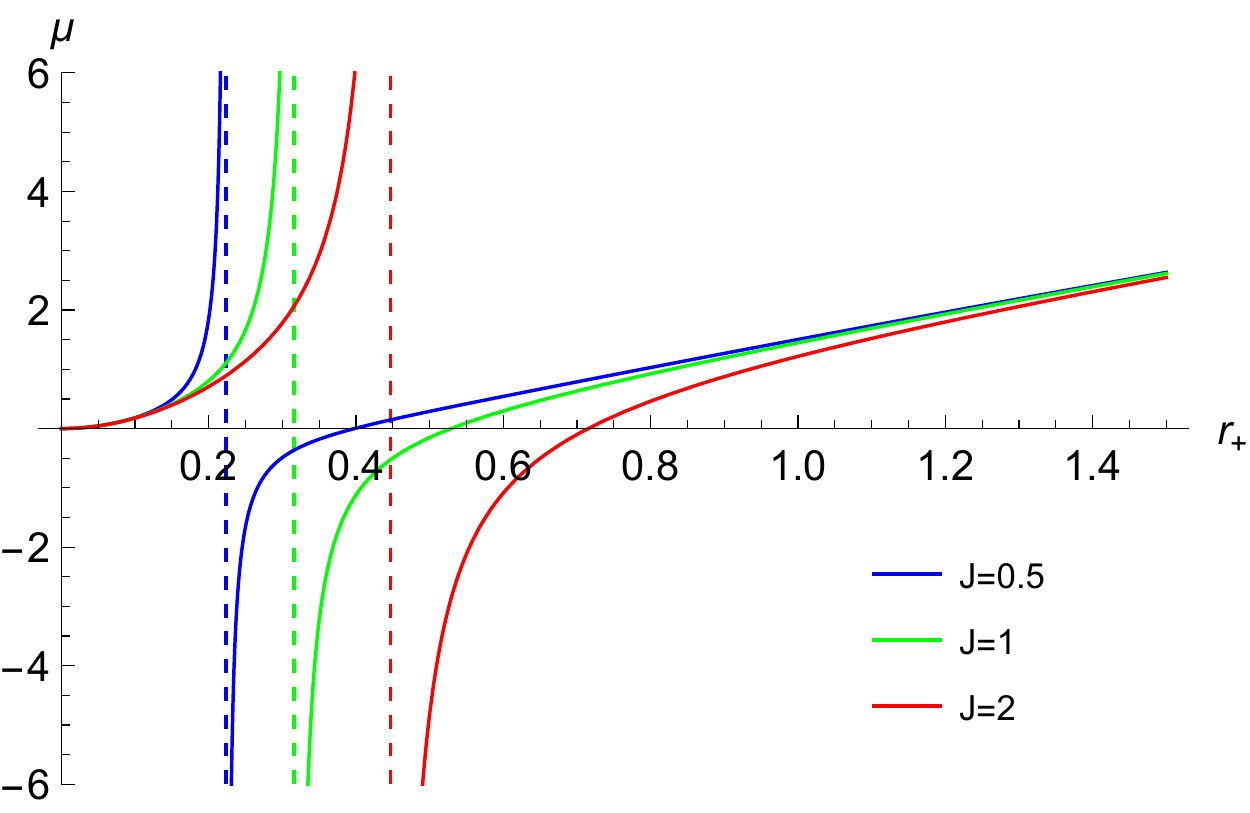}
			\includegraphics[scale=0.48]{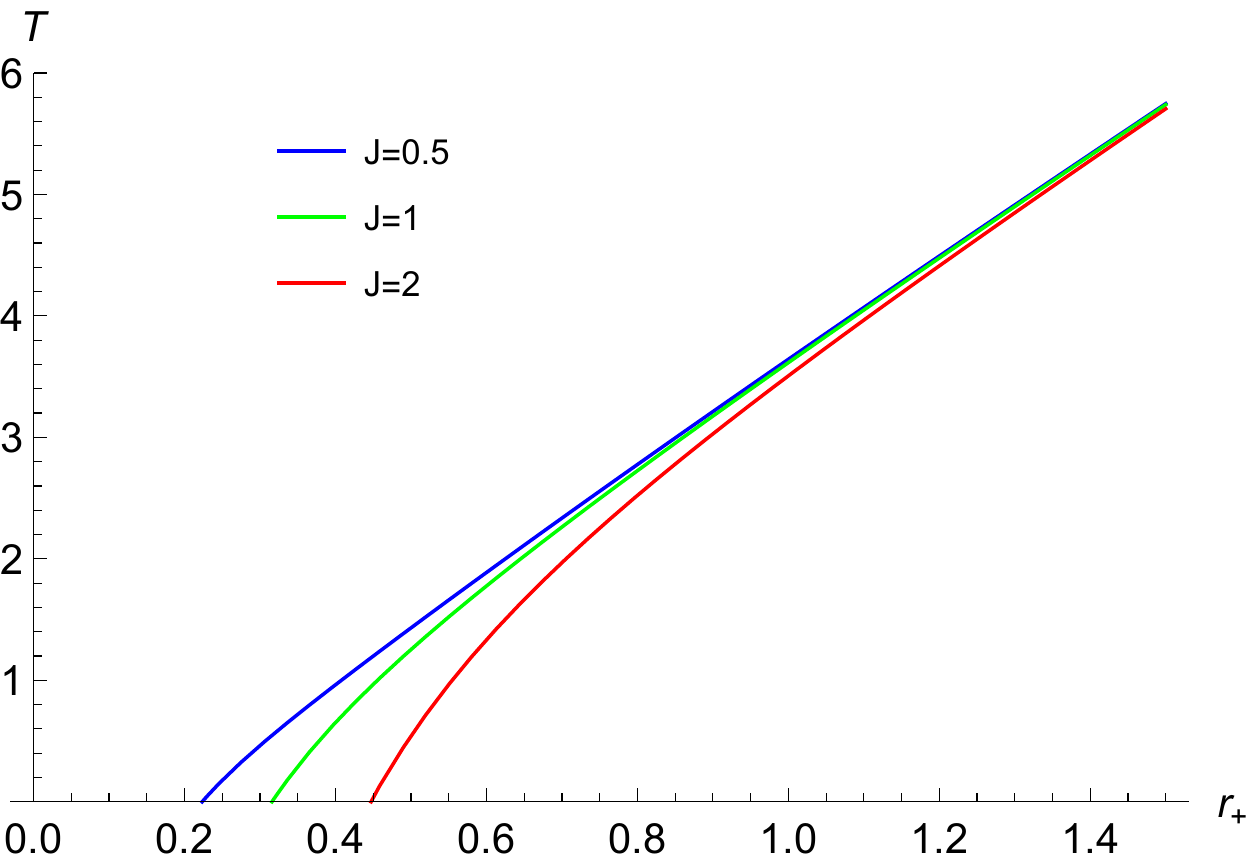}}
		
		\quad
		\subfigure[$\eta=0.5$]{
			\includegraphics[scale=0.54]{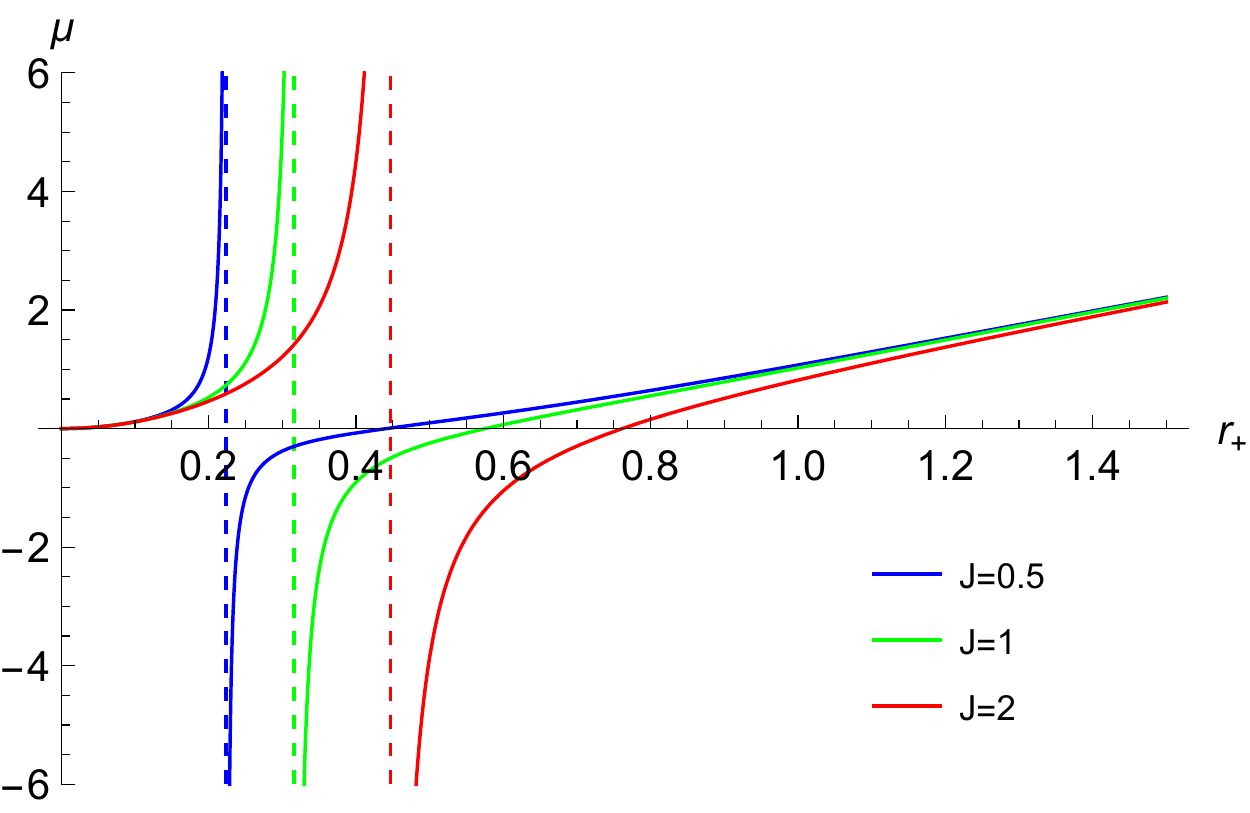}
			\includegraphics[scale=0.48]{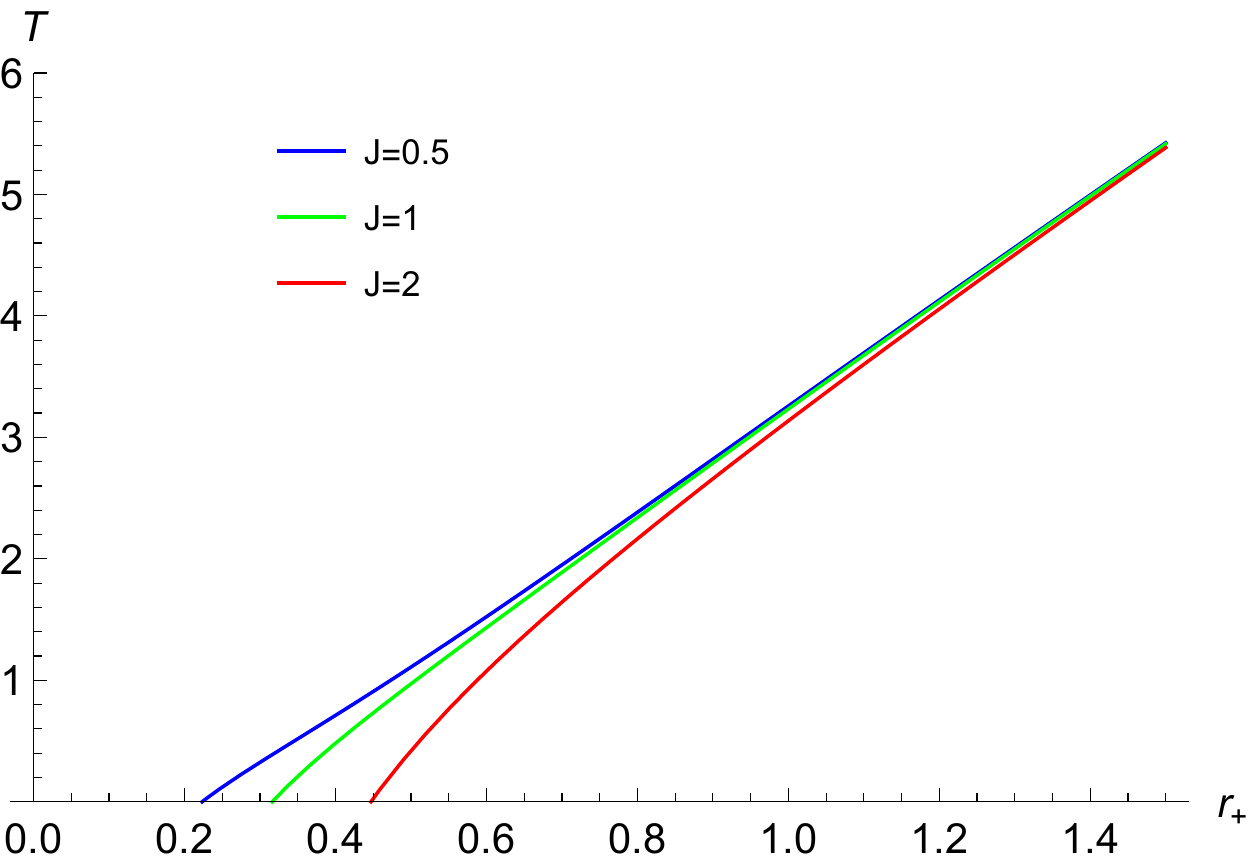}}
		\caption{Plots of the Joule-Thomson coefficient(left) and Hawking temperature $T$(right) in relation to the event horizon $r_+$ with different $\eta$. Here, we set $P=1$ and $J=0.5,1,2$.}
	\end{figure}

	\begin{figure}[htbp]
		\centering
		\subfigure[$\eta=0$]{
			\includegraphics[scale=0.8]{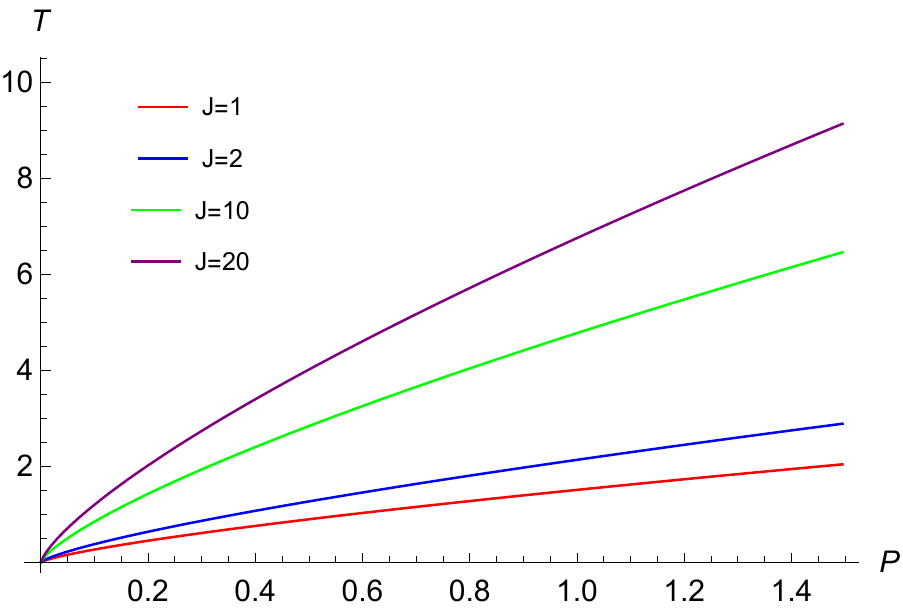}}
		\quad
		\subfigure[$\eta=0.1$]{
			\includegraphics[scale=0.8]{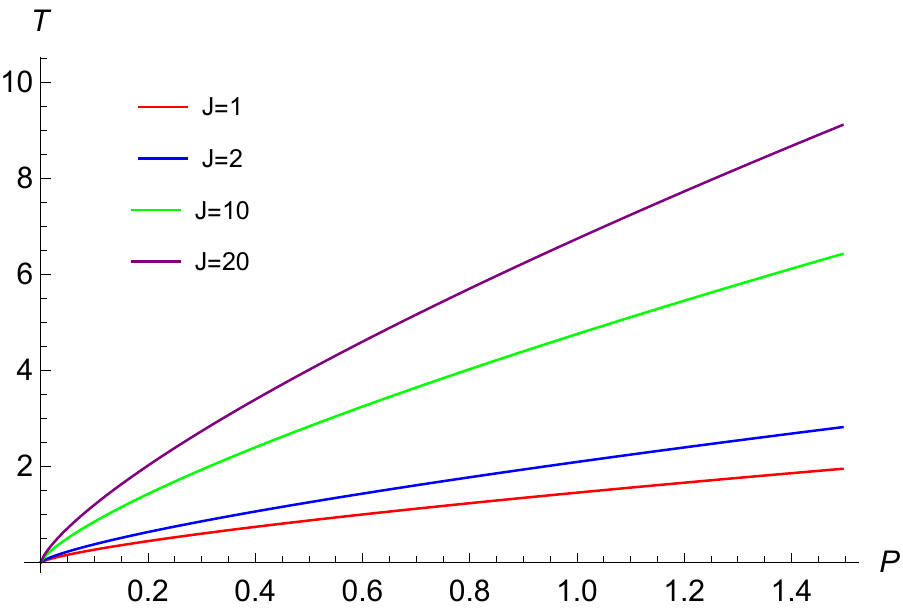}}
		\quad
		\subfigure[$\eta=0.2$]{
			\includegraphics[scale=0.8]{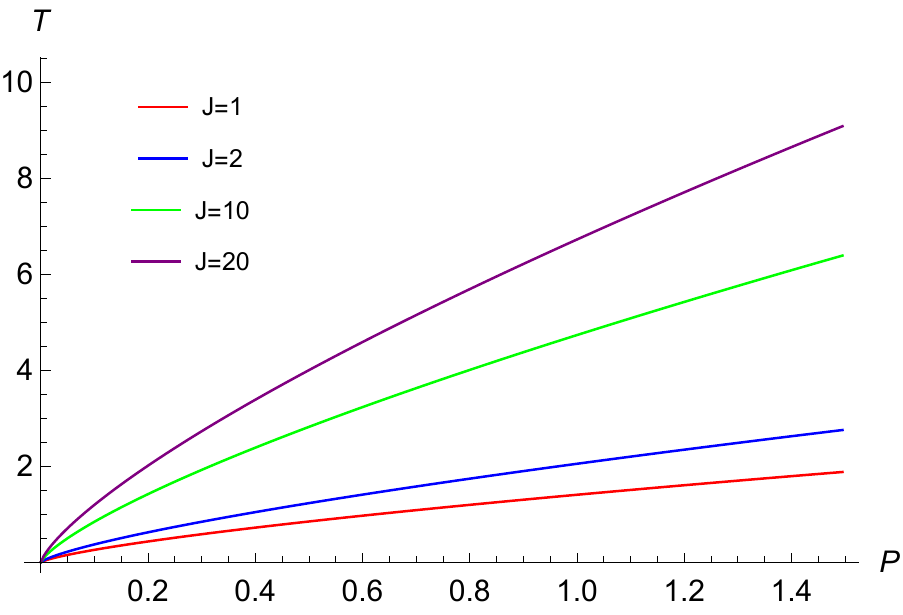}}
		\quad
		\subfigure[$\eta=0.5$]{
			\includegraphics[scale=0.8]{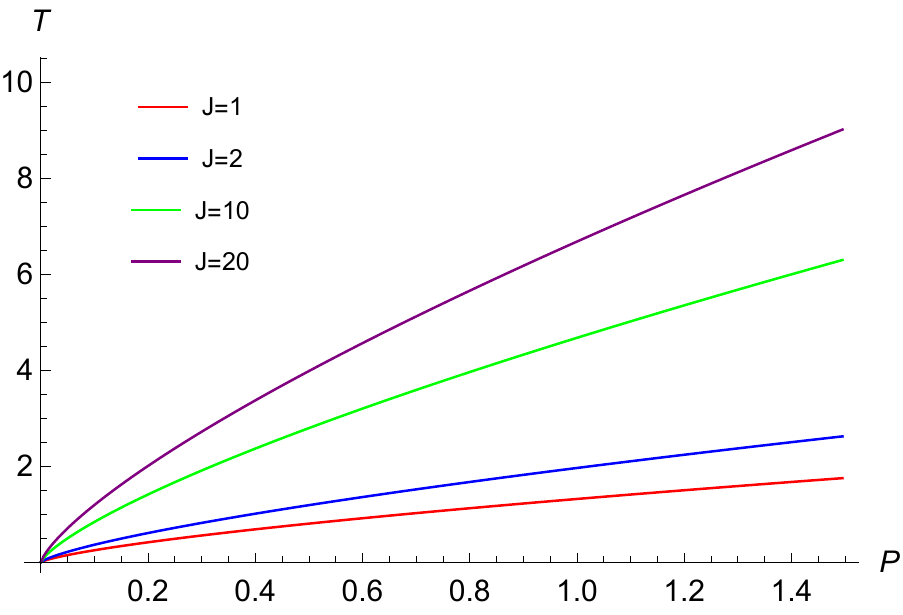}}
		\caption{Plots of inversion curves for a BTZ black hole in rainbow gravity with different $\eta$.}
	\end{figure}

	\begin{figure}[htbp]
		\centering
		\subfigure[$J=1$]{
			\includegraphics[scale=0.86]{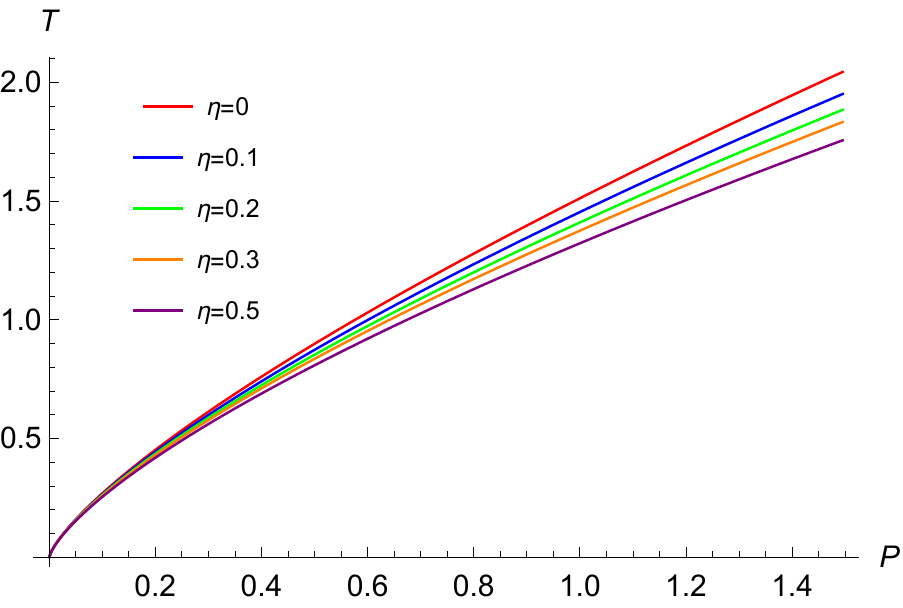}}
		\quad
		\subfigure[$J=2$]{
			\includegraphics[scale=0.85]{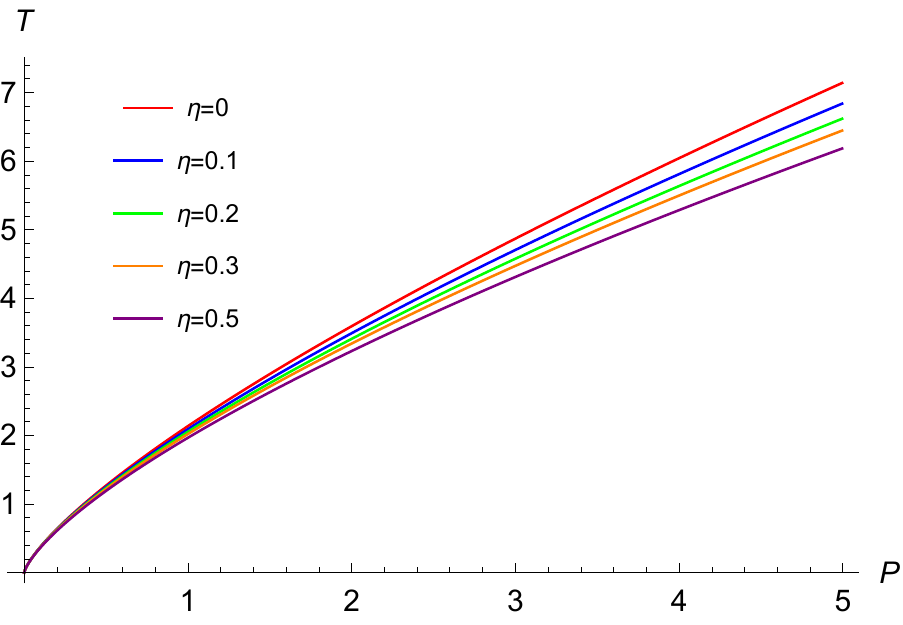}}
		\quad
		\subfigure[$J=10$]{
			\includegraphics[scale=0.86]{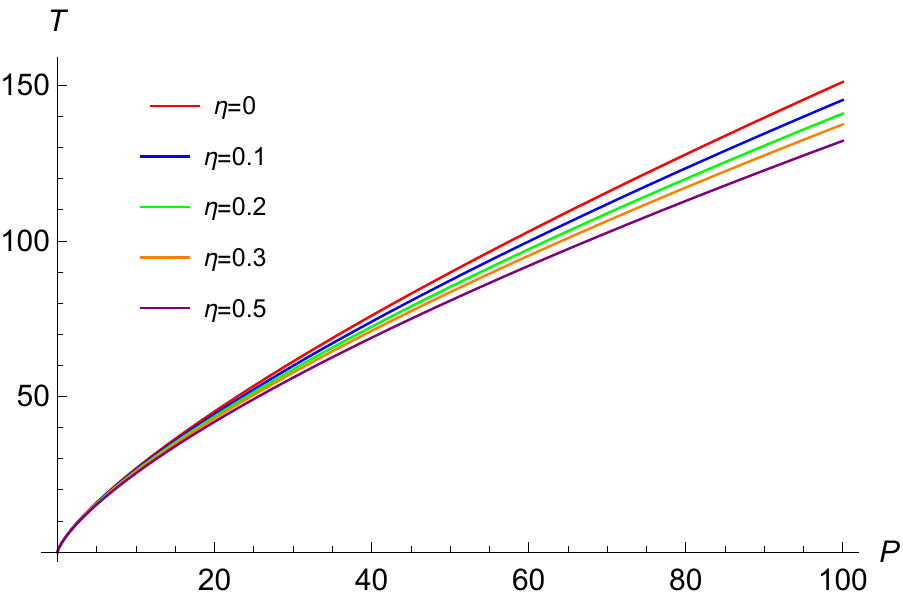}}
		\quad
		\subfigure[$J=20$]{
			\includegraphics[scale=0.85]{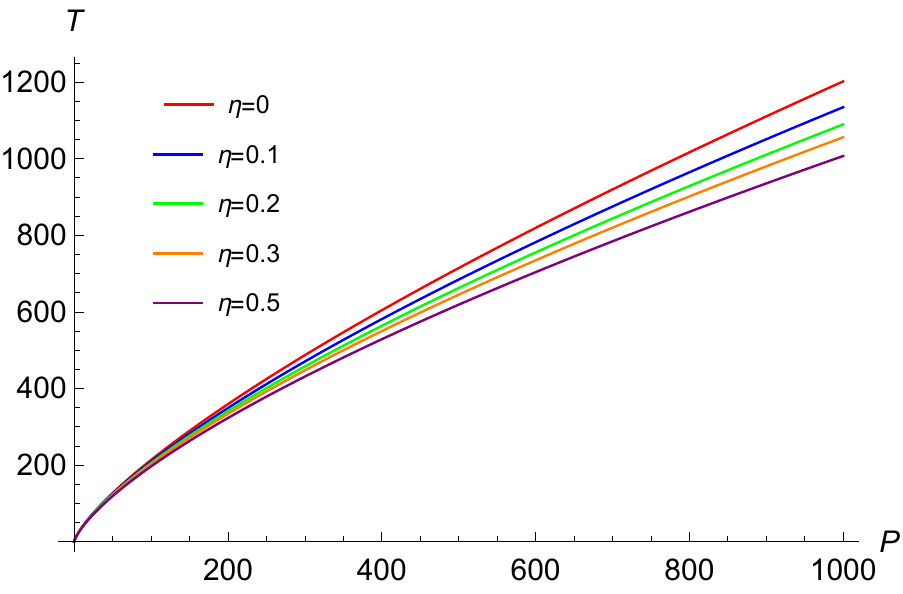}}
		\caption{Plots of inversion curves for a BTZ black hole in rainbow gravity with different $J$.}
	\end{figure}

	Fig.1 shows the Joule–Thomson coefficient $\mu$ and Hawking temperature in relation to the event horizon $r_+$ in rainbow gravity. Here, we fix the pressure $P=1$ and the angular momentum $J$ as 0.5, 1, and 2. For different $\eta$, the divergence point of the Joule-Thomson coefficient exists, remains the same value and corresponds to the zero point of the Hawking temperature, which reveals the relevant information of the extremal black hole. However, as the value of $\eta$ increases, the effects of rainbow gravity are to slow down the speed of the increase of the Joule–Thomson coefficient and make its zero point larger.

	In rainbow gravity, the Hawking temperature of the rotating BTZ black hole can be written as\cite{Mu:2019jjw}
	\begin{equation}\label{4}
		\begin{aligned}
			T&=\left(\frac{r_{+}}{2 \pi l^2}-\frac{J^2}{8 \pi r_{+}^3} \right) \sqrt{1-\eta \frac{  \left(m^2 r_{+}^2+1\right)}{\eta +r_{+}^2}}\\
			&=\left(4 P r_{+}-\frac{J^2}{8 \pi  r_{+}^3}\right) \sqrt{1-\eta \frac{  \left(m^2 r_{+}^2+1\right)}{\eta +r_{+}^2}}.
		\end{aligned}
	\end{equation}
	Then, the inversion temperature can be obtained as follow
	\begin{equation}\label{5}
		T_i= V \left(  \frac{\partial T}{\partial V}  \right)_{P} = \frac{\sqrt{\frac{r_{+}^2 \left(1-\eta  m^2\right)}{\eta +r_{+}^2}} \left(J^2 \left(2 \eta +3 r_{+}^2\right)+32 \pi  P r_{+}^4 \left(2 \eta +r_{+}^2\right)\right)}{16 \pi  r_{+}^3 \left(\eta +r_{+}^2\right)}.
	\end{equation}
	Subtracting equation(\ref{5}) from equation(\ref{4}) yields
	\begin{equation}
		\frac{\sqrt{\frac{r_{+}^2 \left(1-\eta  m^2\right)}{\eta +r_{+}^2}} \left(-J^2 \left(4 \eta +5 r_{+}^2\right)+32 \pi  P_{i} r_{+}^6\right)}{16 \pi  r_{+}^3 \left(\eta +r_{+}^2\right)}=0.
	\end{equation}
	Solving this equation for $r_+$, the positive and real root is
	\begin{equation}
		r_{+}=\frac{\sqrt{4 M + \sqrt{16 M^2-2 \pi  J^2 P_{i}}}}{2 \sqrt{2 \pi  P_{i}}}.
	\end{equation}
	Substituting this root into equation(\ref{5}) at $P = P_i$, we can get the inversion temperature with parameters $J$ and $P_i$. Set $P_i=0$, and the minimum of inversion temperature is given by
	\begin{equation}
		T_{i}^{min}=0,
	\end{equation}
	which means the black hole becomes an extremal black hole.
	
	A rotating BTZ black hole does not have $P-V$ critical behavior, which means that $P_c$, $T_c$, and $V_c$ do not exist. Therefore, the black hole is always thermodynamically stable, and the ratio between the minimum inversion temperature $T_i^{min}$ and $T_c$ also do not exist. This is different from some other black holes. To better investigate the effects of rainbow gravity acting on the Joule-Thomson expansion in a rotating BTZ black hole, the inversion curves of the black hole with different $\eta$ are shown in Fig.2 and 3, and the isoenthalpic curves are shown in Fig.4, 5, 6 and 7.
	
	From Fig.2, it is obvious that the inversion temperature increases monotonically with the increase of the inversion pressure, while the slope of the curve decreases with the increase of the inversion pressure. Meanwhile, the inversion curves are not closed, which means it does not terminate at any point. This is similar to the result of rotating BTZ black holes without rainbow gravity. From Fig.3, we can find that the inversion temperature drops with the increase of $\eta$. And when $J$ becomes lager, the effect of rainbow gravity to the inversion temperature becomes smaller.
	
	\begin{figure}[htbp]
		\centering
		\subfigure[$\eta=0$]{
			\includegraphics[scale=0.45]{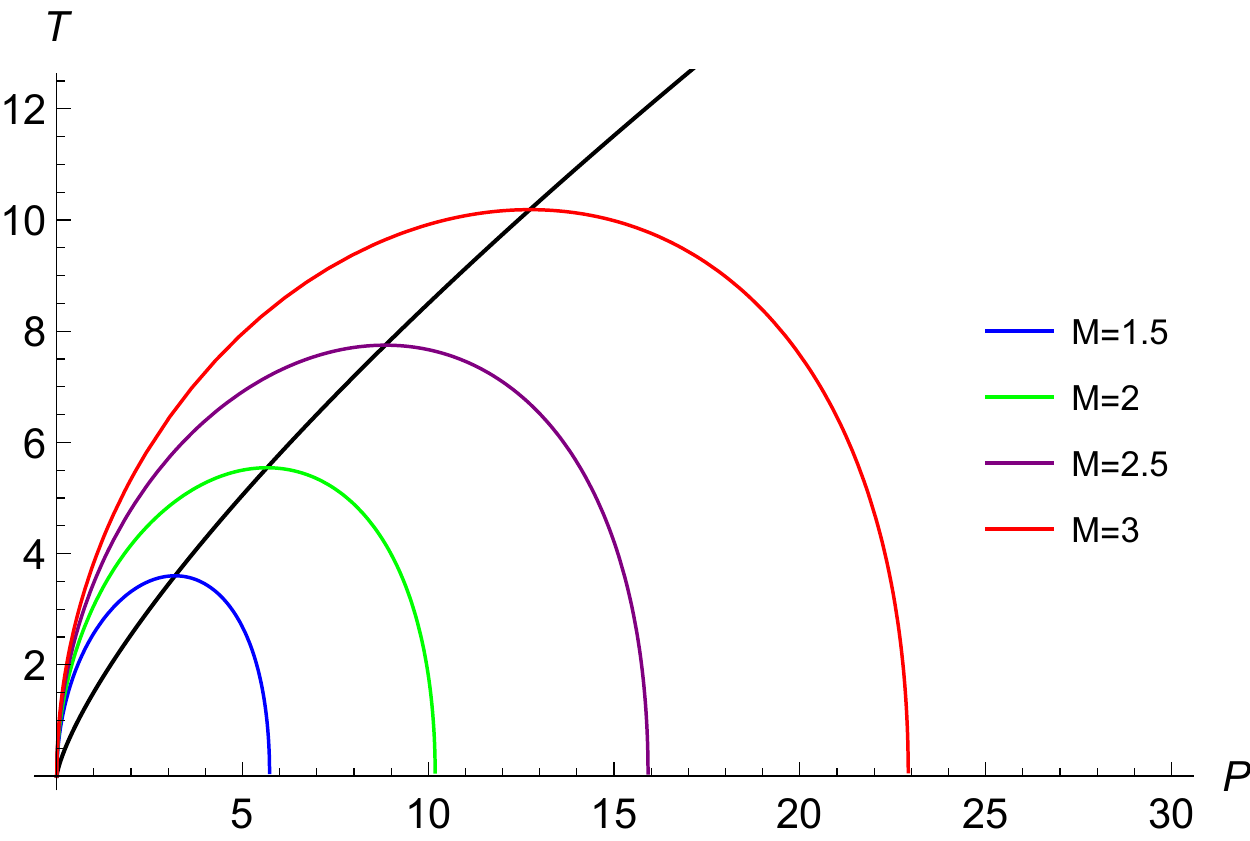}}
		\quad
		\subfigure[$\eta=0.1$]{
			\includegraphics[scale=0.45]{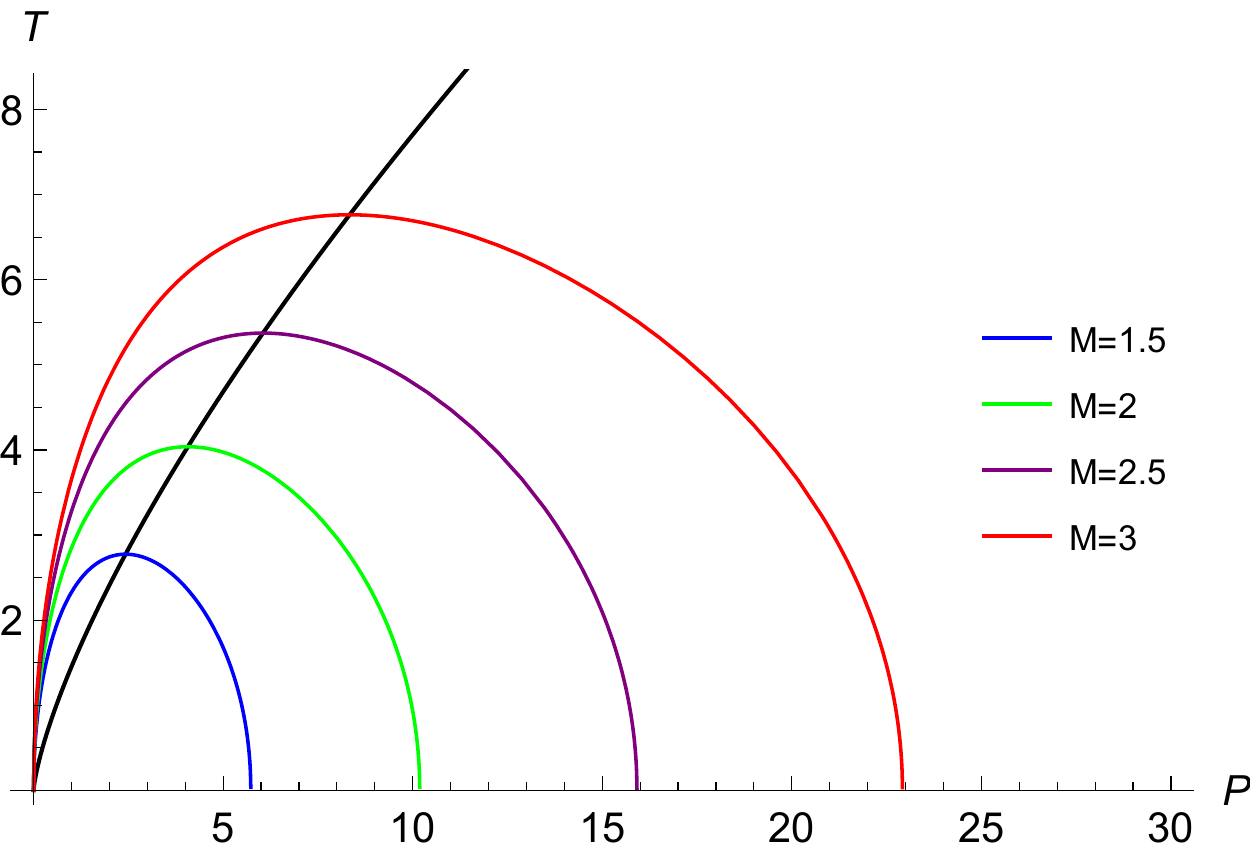}}
		\quad
		\subfigure[$\eta=0.2$]{
			\includegraphics[scale=0.45]{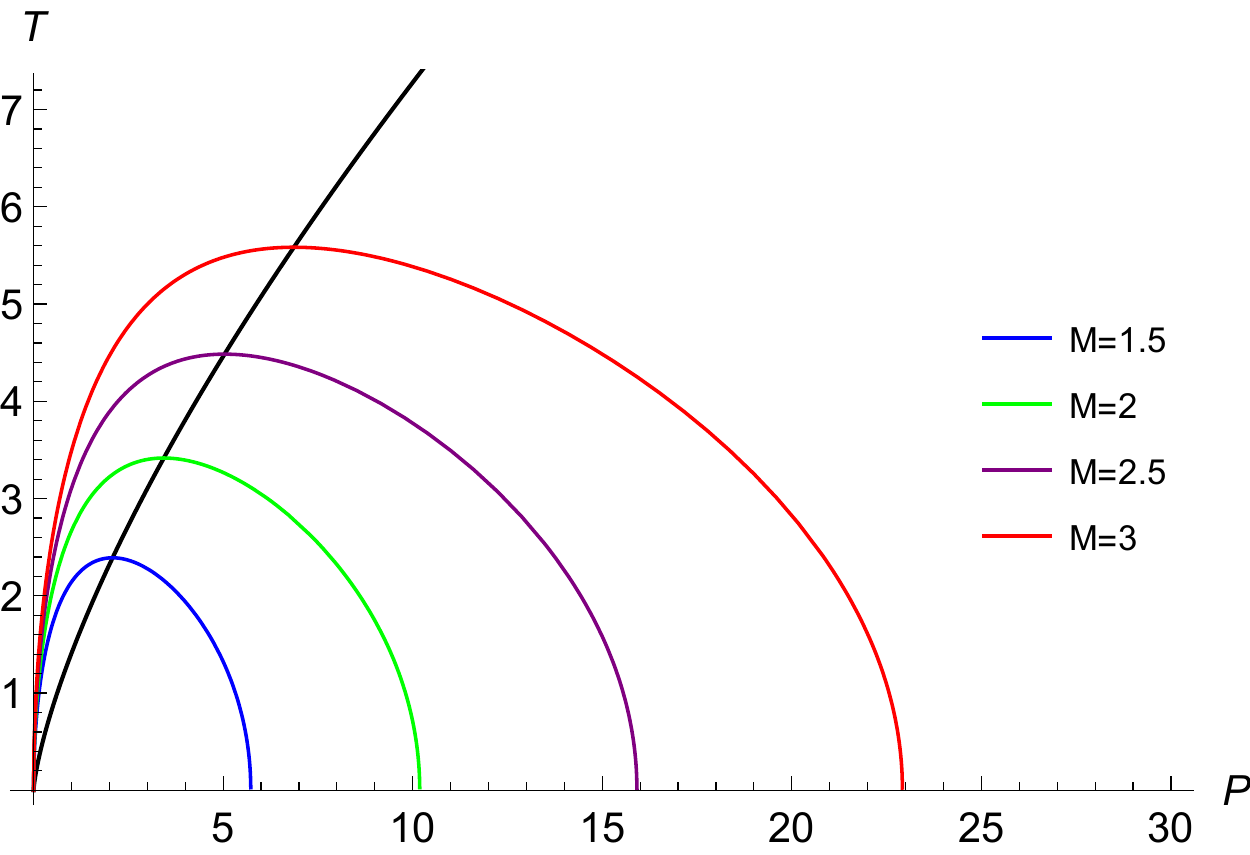}}
		\quad
		\subfigure[$\eta=0.5$]{
			\includegraphics[scale=0.45]{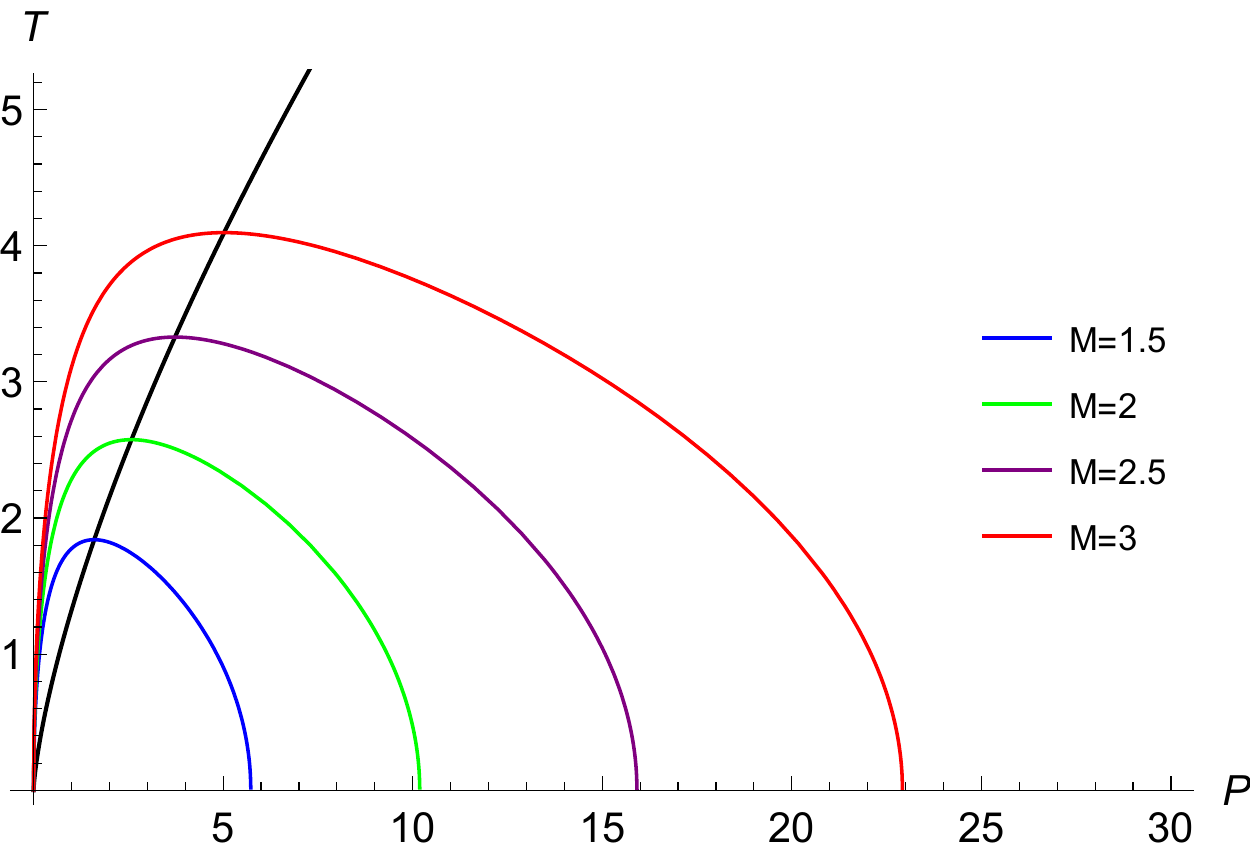}}
		\caption{Plots of inversion and isenthalpic curves for a BTZ black holein rainbow gravity with $J=1$. The black lines are the inversion curves.}
	\end{figure}
	
	\begin{figure}[htbp]
		\centering
		\subfigure[$\eta=0$]{
			\includegraphics[scale=0.45]{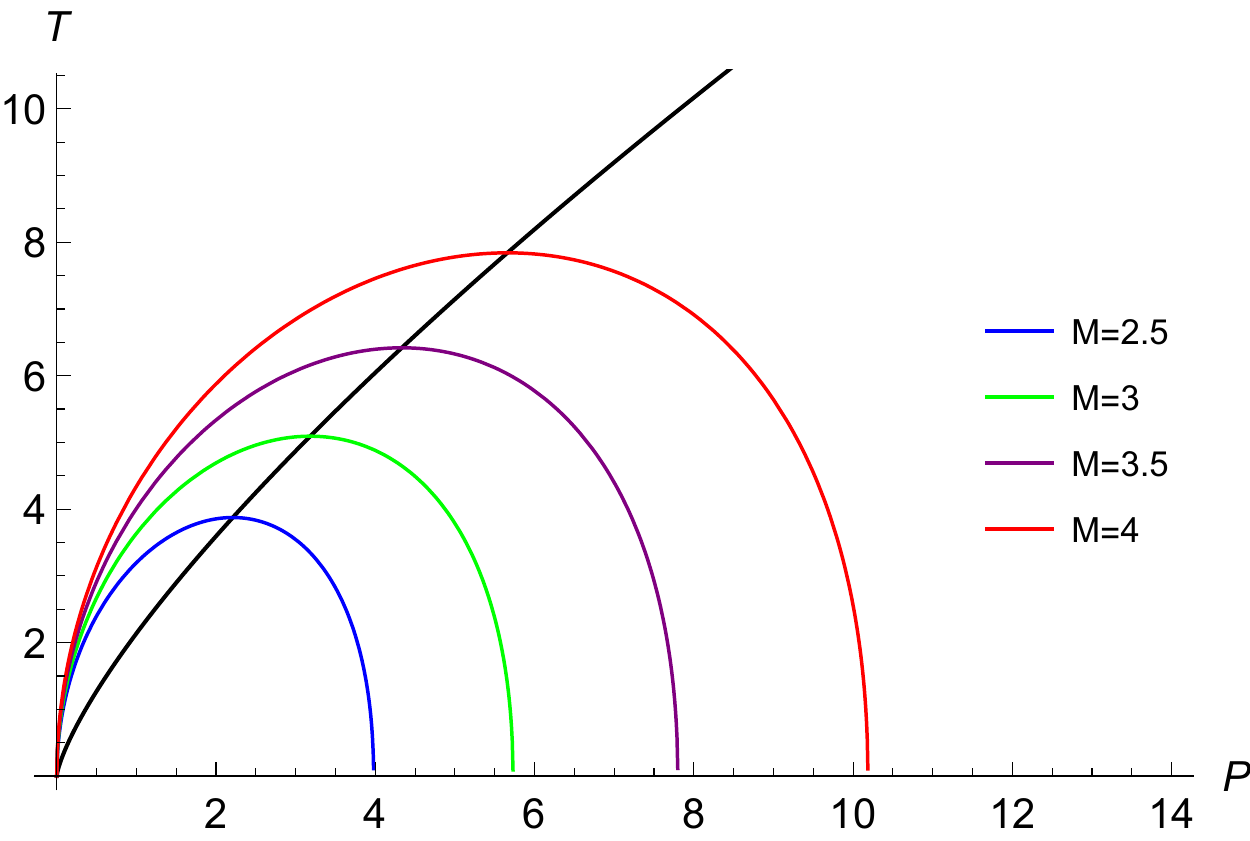}}
		\quad
		\subfigure[$\eta=0.1$]{
			\includegraphics[scale=0.45]{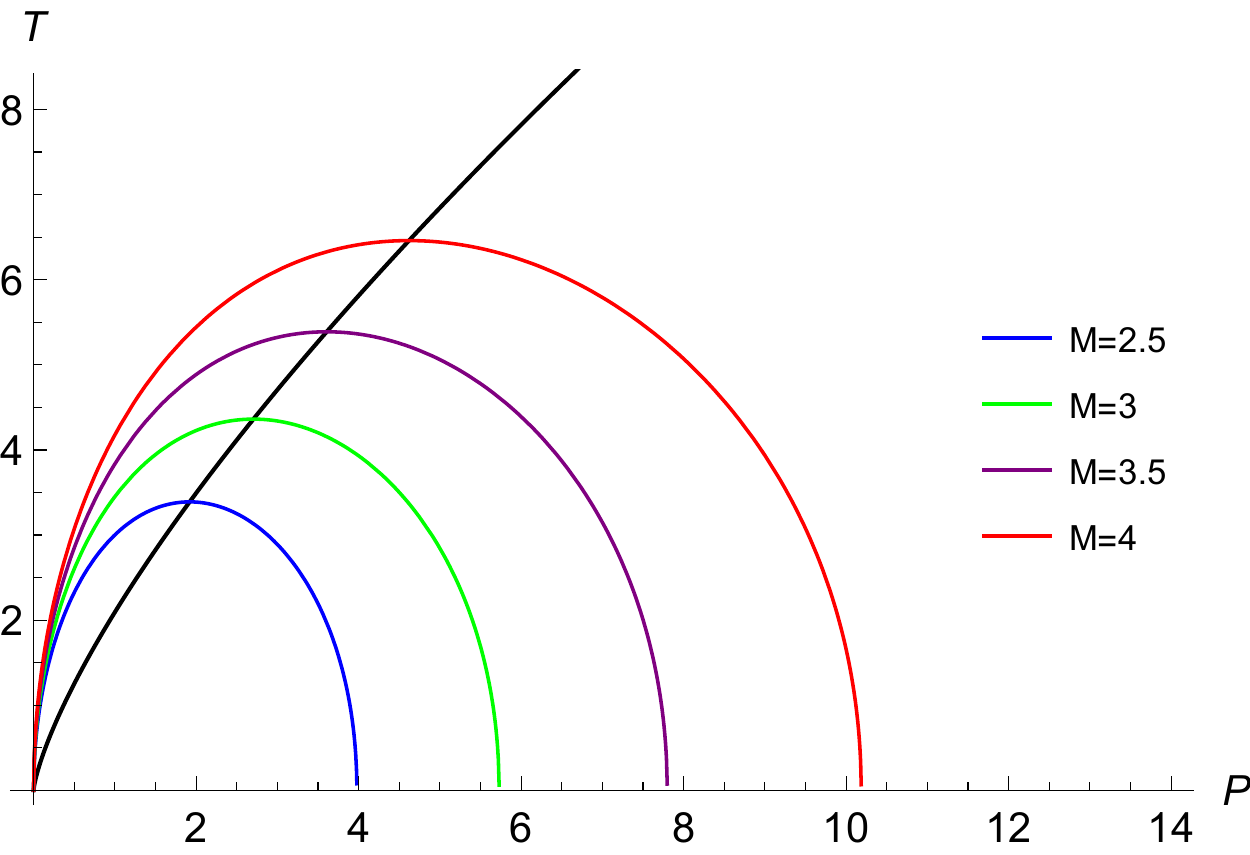}}
		\quad
		\subfigure[$\eta=0.2$]{
			\includegraphics[scale=0.45]{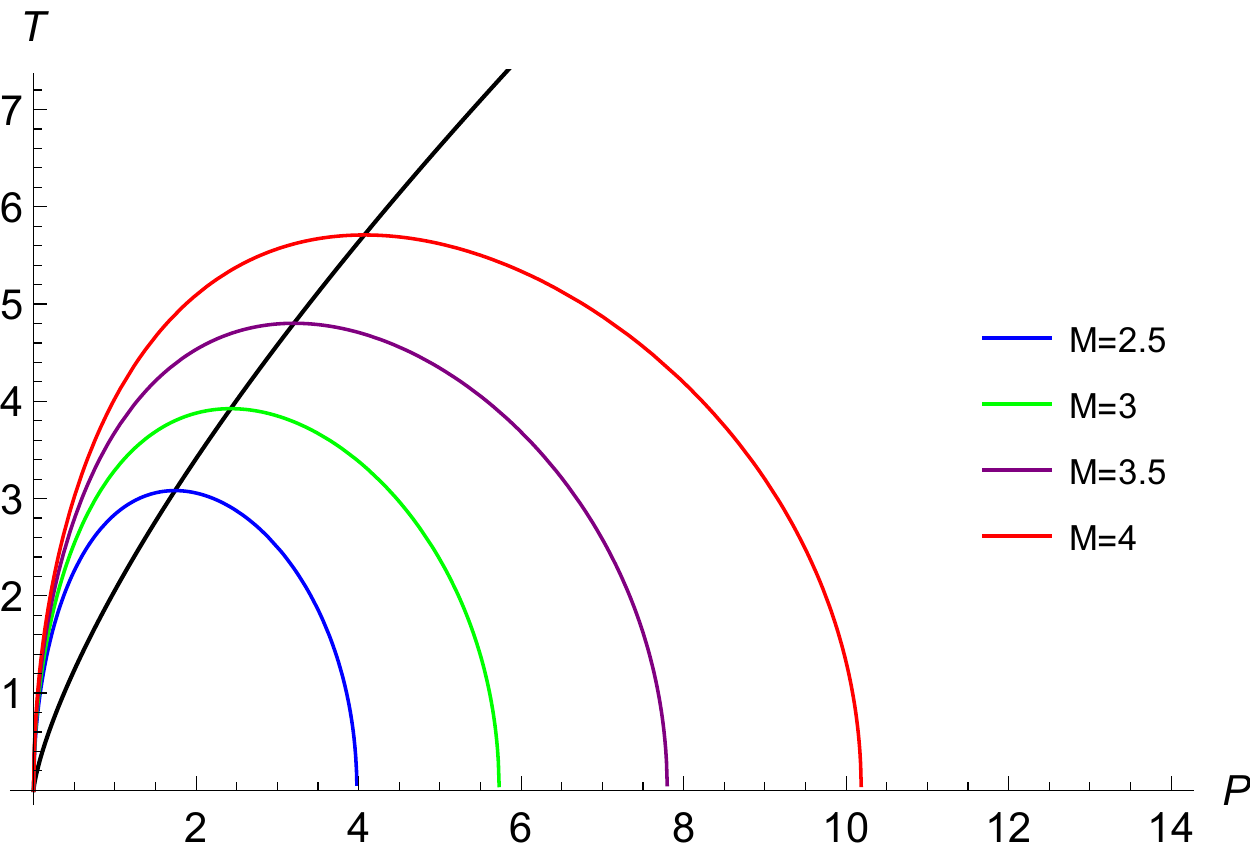}}
		\quad
		\subfigure[$\eta=0.5$]{
			\includegraphics[scale=0.45]{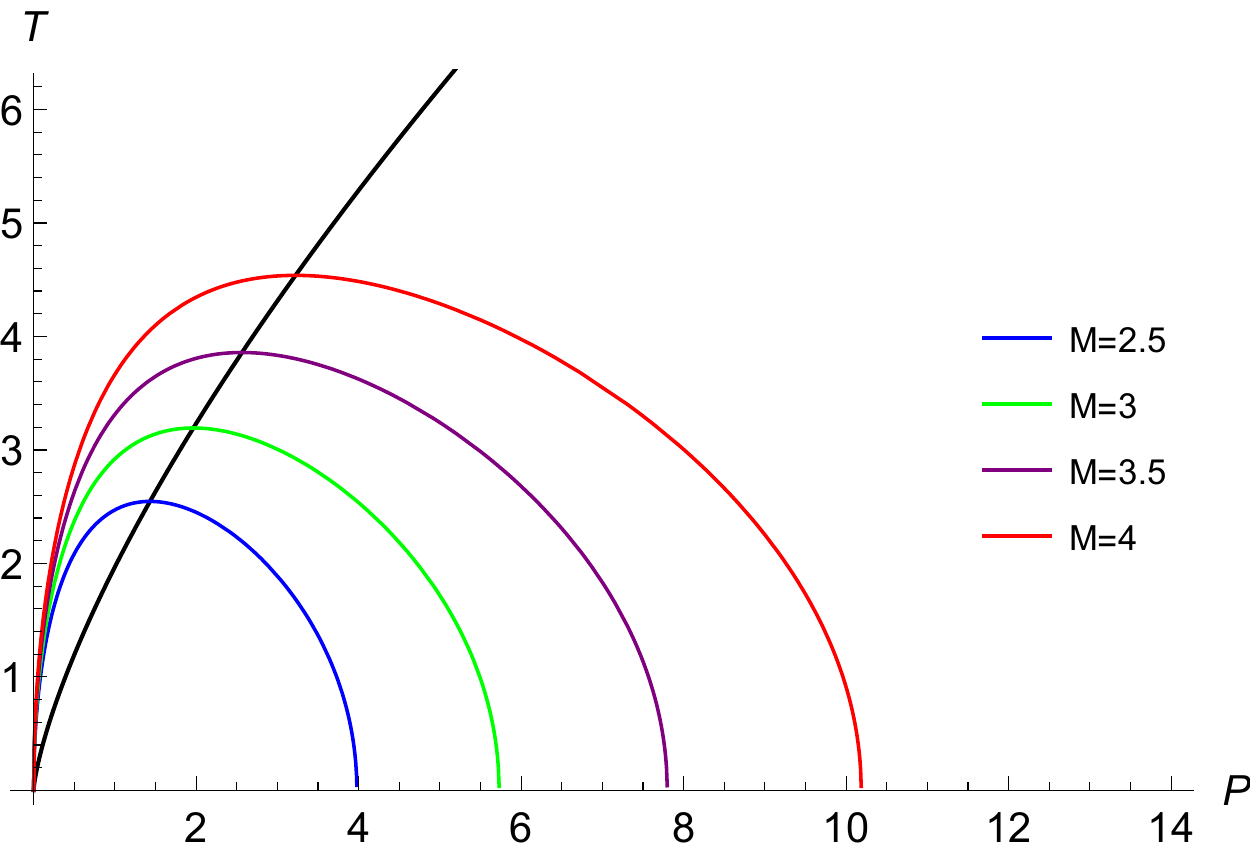}}
		\caption{Plots of inversion and isenthalpic curves for a BTZ black holein rainbow gravity with $J=2$. The black lines are the inversion curves.}
	\end{figure}
	
	\begin{figure}[htbp]
		\centering
		\subfigure[$\eta=0$]{
			\includegraphics[scale=0.45]{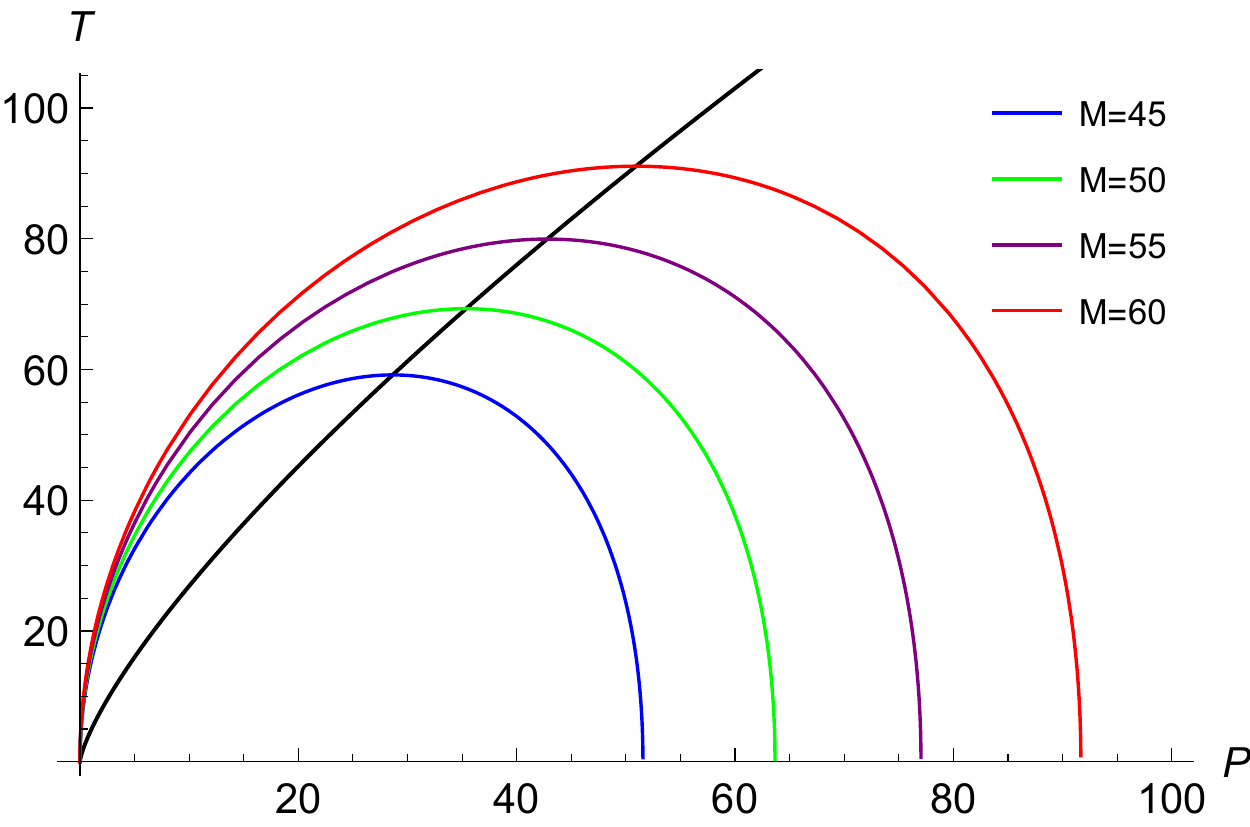}}
		\quad
		\subfigure[$\eta=0.1$]{
			\includegraphics[scale=0.45]{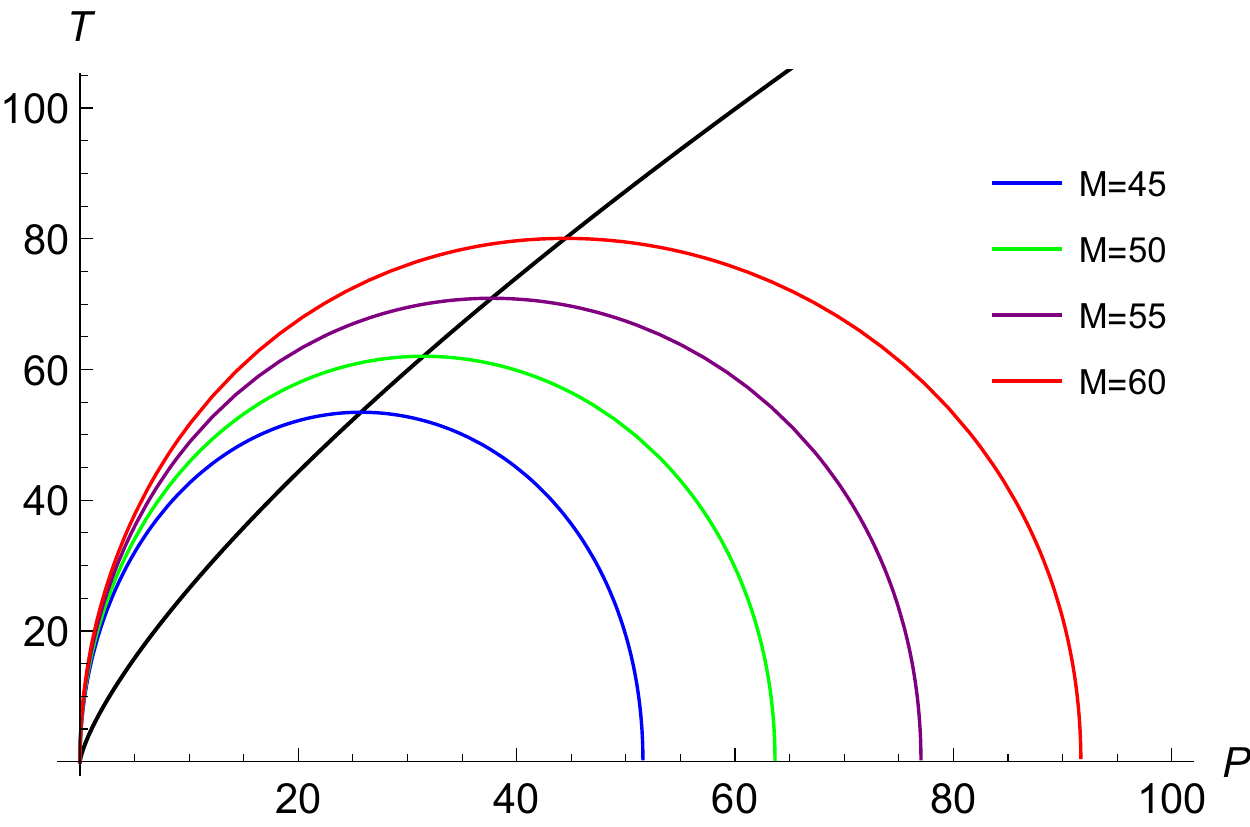}}
		\quad
		\subfigure[$\eta=0.2$]{
			\includegraphics[scale=0.45]{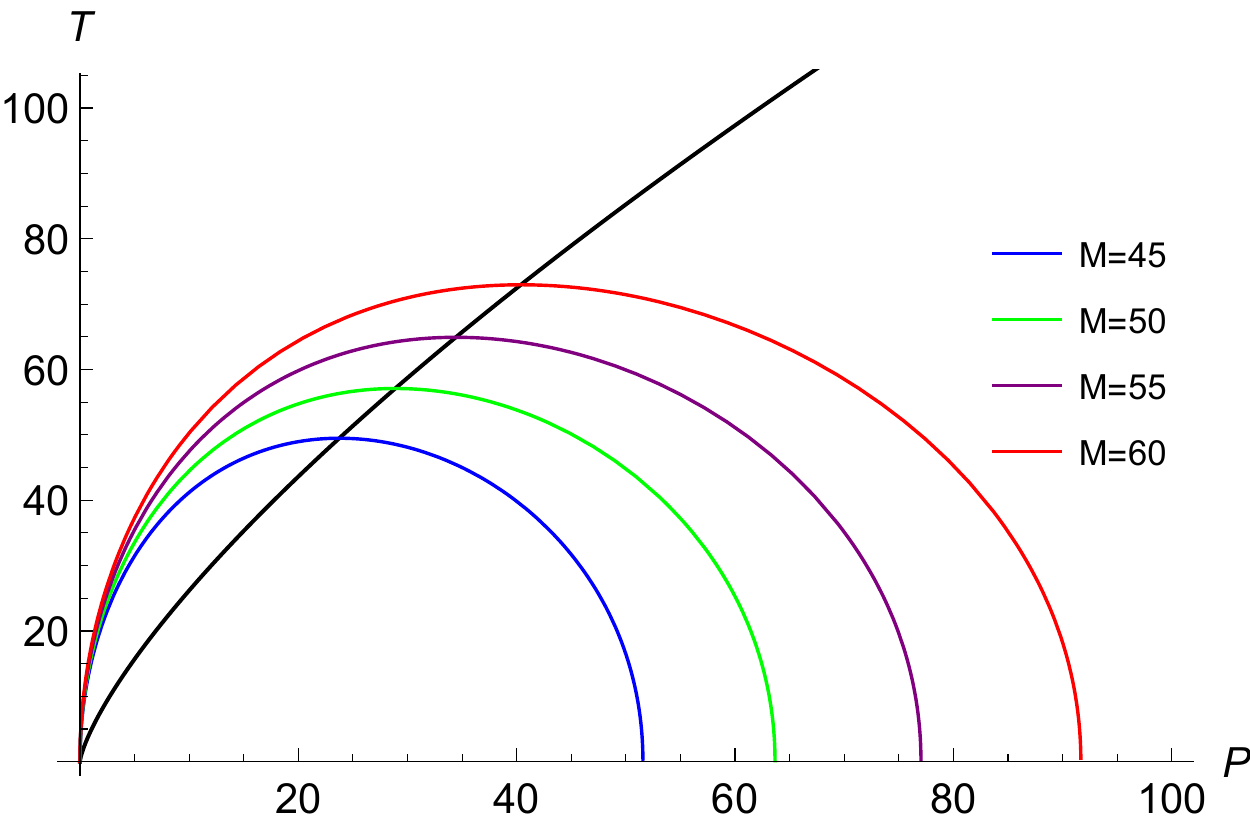}}
		\quad
		\subfigure[$\eta=0.5$]{
			\includegraphics[scale=0.45]{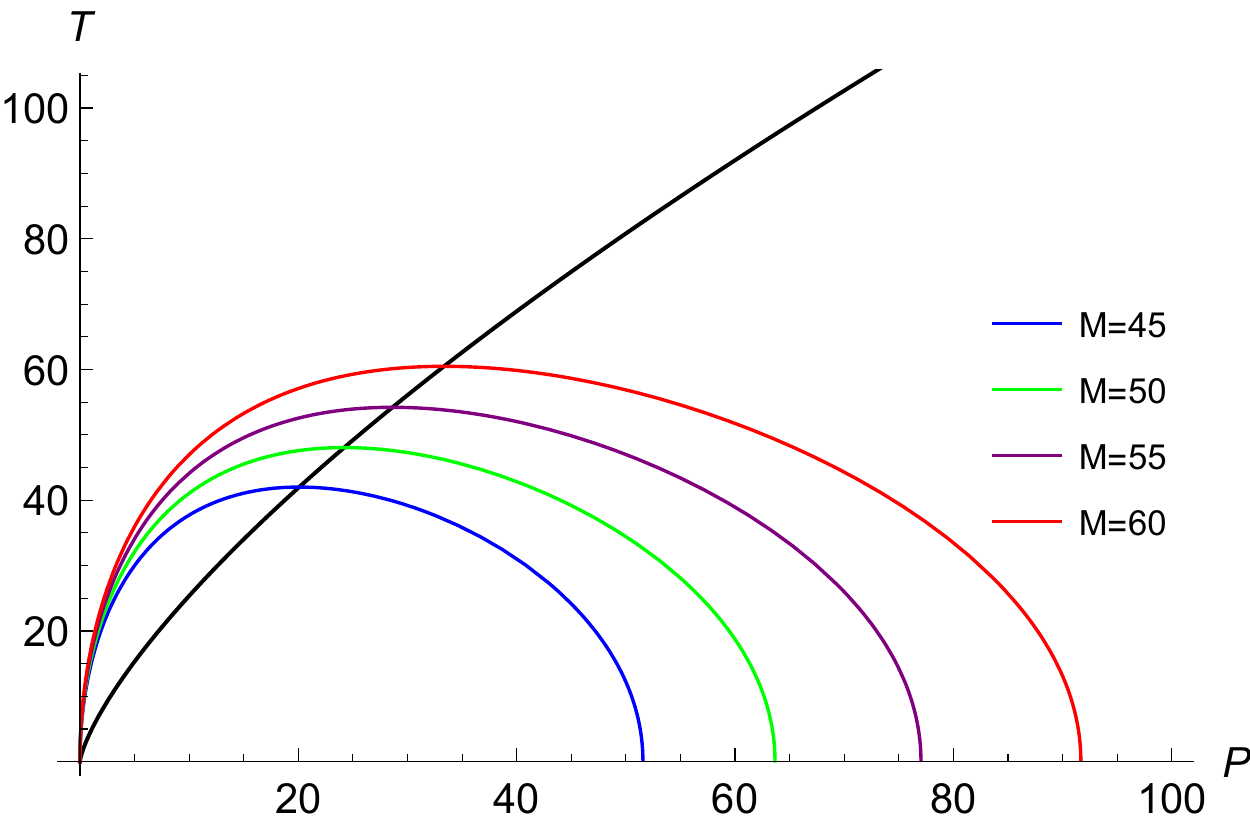}}
		\caption{Plots of inversion and isenthalpic curves for a BTZ black holein rainbow gravity with $J=10$. The black lines are the inversion curves.}
	\end{figure}
	
	\begin{figure}[htbp]
		\centering
		\subfigure[$\eta=0$]{
			\includegraphics[scale=0.45]{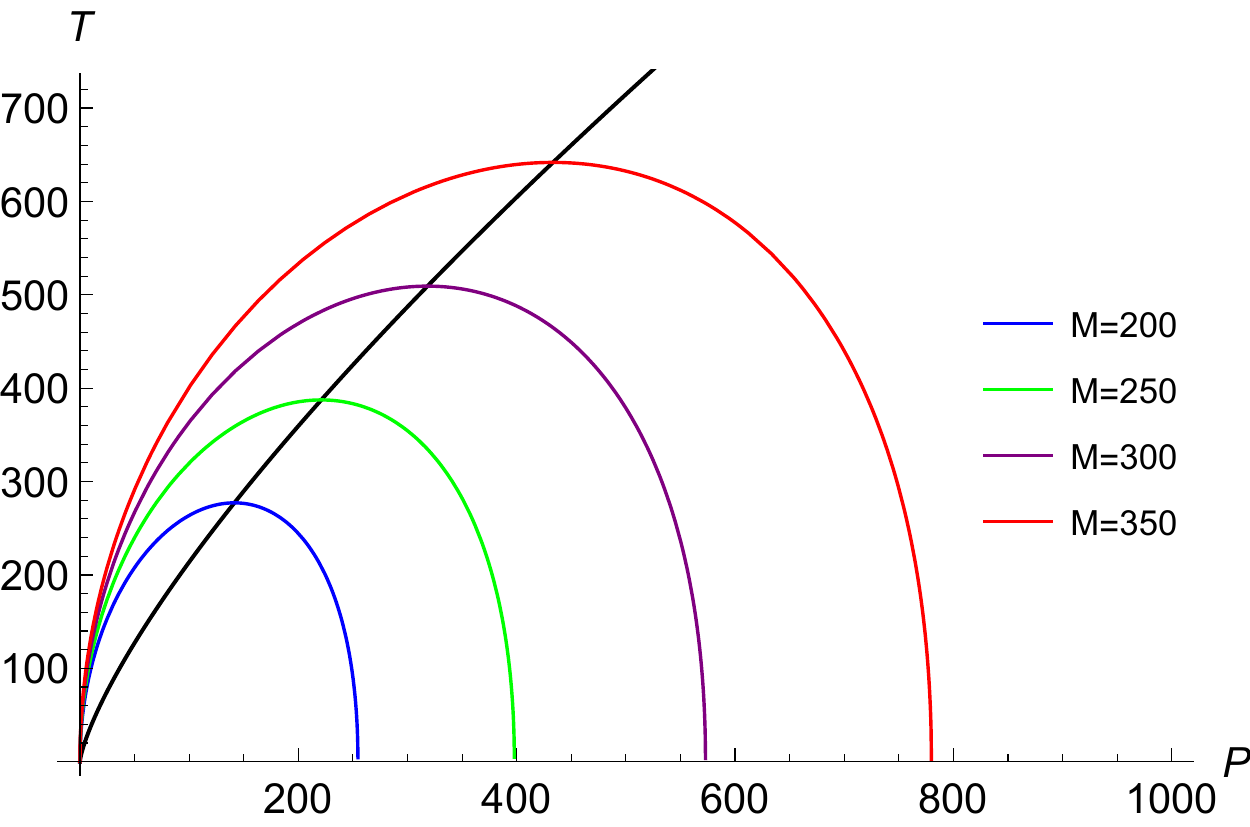}}
		\quad
		\subfigure[$\eta=0.1$]{
			\includegraphics[scale=0.45]{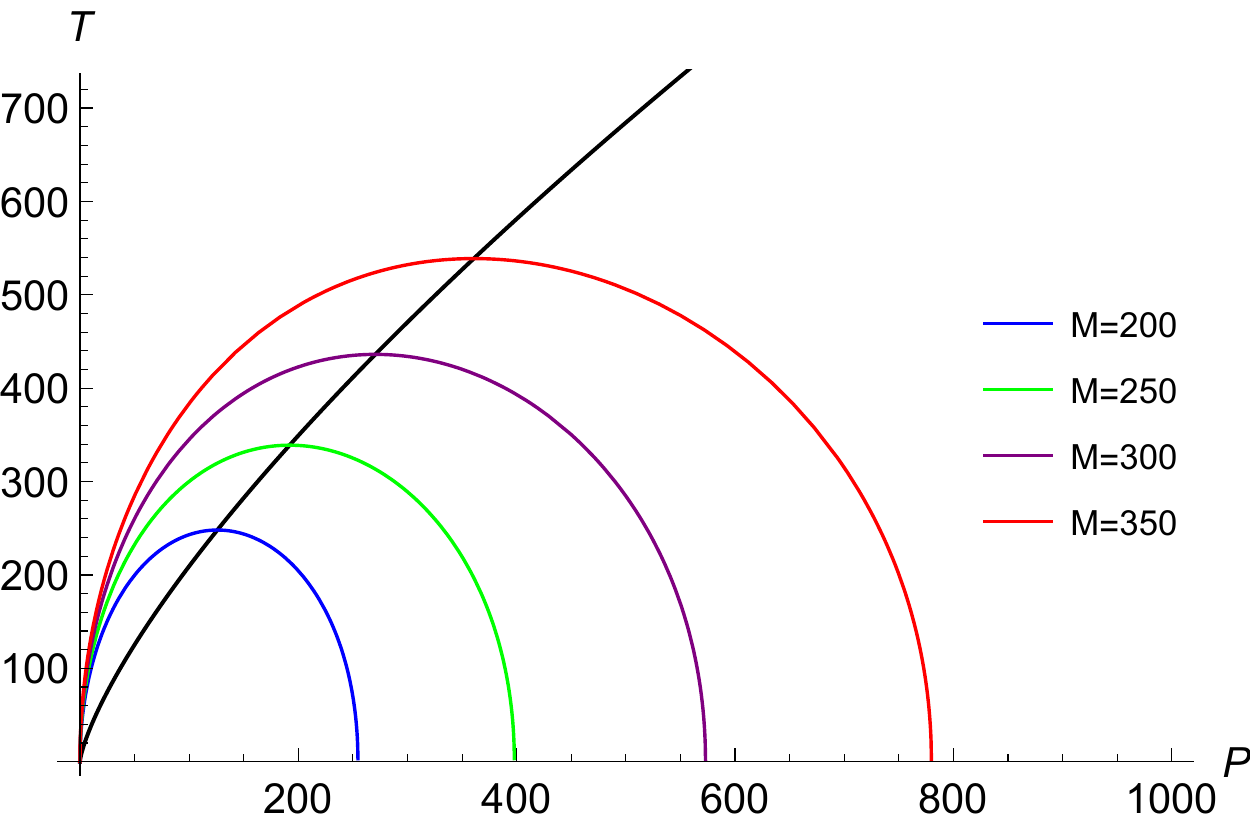}}
		\quad
		\subfigure[$\eta=0.2$]{
			\includegraphics[scale=0.45]{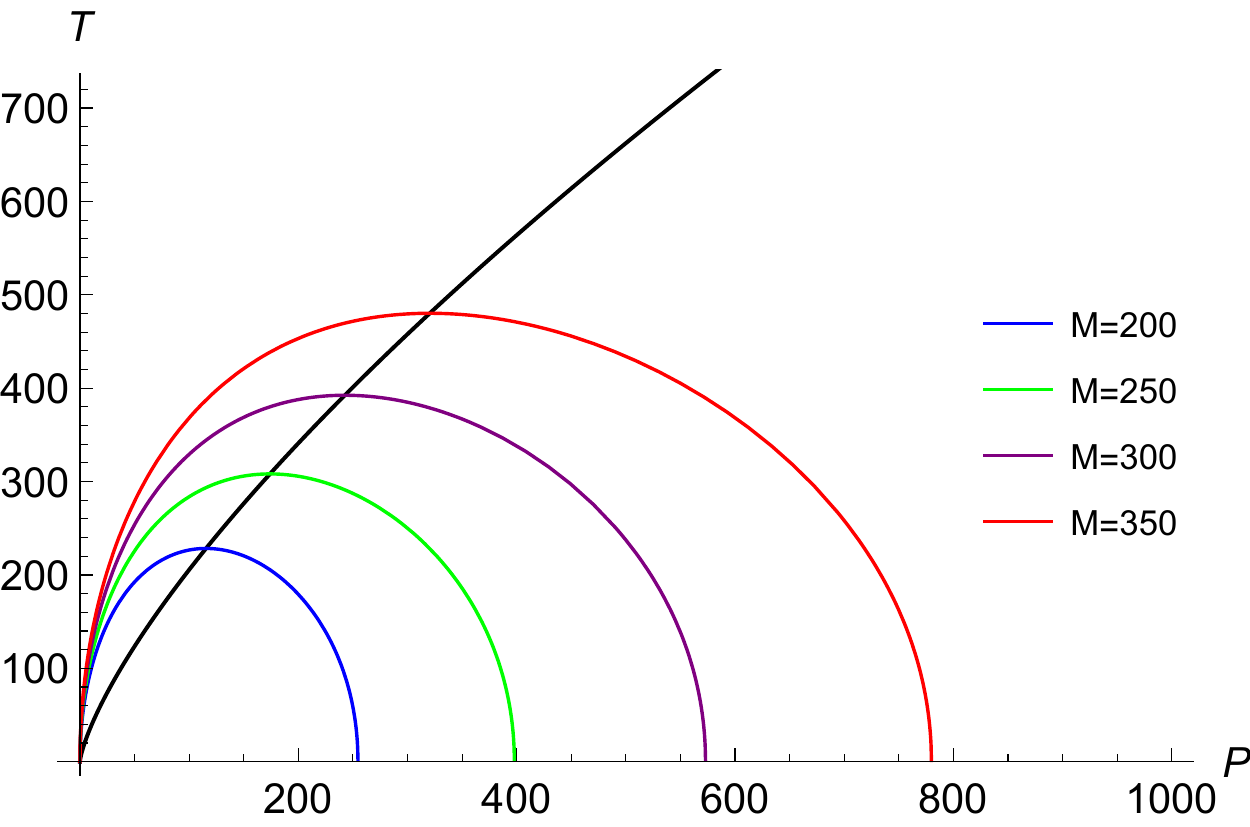}}
		\quad
		\subfigure[$\eta=0.5$]{
			\includegraphics[scale=0.45]{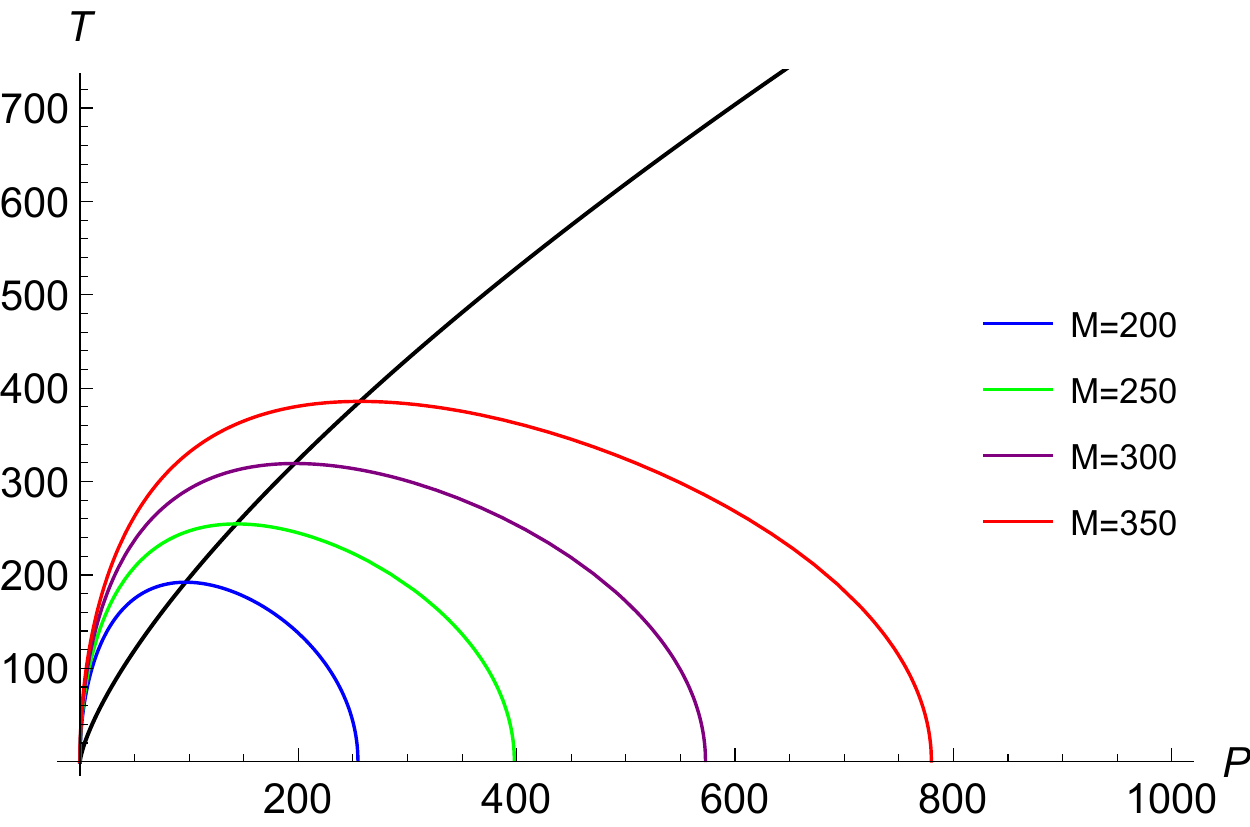}}
		\caption{Plots of inversion and isenthalpic curves for a BTZ black holein rainbow gravity with $J=20$. The black lines are the inversion curves.}
	\end{figure}

	\begin{table}[htbp]
		\centering
		\caption{Existence of critical behavior and the ratio of the minimum inversion in Van der Waals fluid and various black holes.}
		
		\begin{tabular}{p{4.5cm}| p{3.5cm} p{2cm} p{2cm} p{2cm} p{2cm} }
			\hline\hline
			Type &Critical behavior &$T_i^{min}$ &$T_c$ &Ratio &Literature \\
			\hline
			
			Van der Waals fluid &Exist &Exist &Exist &0.75 &\cite{Okcu:2016tgt} \\
			
			Kerr-AdS BH &Exist &Exist &Exist &0.504622 &\cite{Okcu:2017qgo} \\
			
			$d$-dimensional AdS BH &Exist &Exist &Exist &<0.5 &\cite{Mo:2018rgq} \\
			
			Gauss-Bonnet AdS BH &Exist &Exist &Exist &0.4765 &\cite{Lan:2018nnp} \\
			
			Toruslike BH &Not exist &Exist &Not exist &Not exist &\cite{Liang:2021elg} \\
			
			RN-AdS BH &Exist &Exist &Exist &0.5 &\cite{Okcu:2016tgt} \\
			
			Quintessence RN-AdS BH &Exist &Exist &Exist &0.5 &\cite{Ghaffarnejad:2018exz} \\
			
			Cloud of strings RN-AdS BH &Exist &Exist &Exist &0.5 &\cite{Yin:2021akt} \\
			
			Cloud of strings and quintessence-RN-AdS BH &Exist &Exist &Exist &<0.5 &\cite{Yin:2021akt} \\
			
			f(r)-gravity AdS BH &Exist &Exist &Exist &0.5 &\cite{Belhaj:2018cdp} \\
			
			Global-monopole AdS BH &Exist &Exist &Exist &0.5 &\cite{RizwanCL:2018cyb} \\
			
			Bardeen AdS BH &Exist &Exist &Exist &0.536622 &\cite{Pu:2019bxf} \\
			
			Born–Infeld AdS BH &Exist &Exist &Exist &$\approx$0.5 &\cite{Bi:2020vcg} \\
			
			Rotating BTZ BH &Not exist &Exist &Not exist &Not exist &\cite{Liang:2021xny} \\
			
			Rainbow charged AdS BH &Exist &Exist &Exist &$\approx$0.5 &\cite{MahdavianYekta:2019dwf}  \\
			
			Rainbow rotating BTZ BH &Not exist &Exist &Not exist &Not exist & \\
			
			\hline\hline
		\end{tabular}
	\end{table}

	Joule-Thomson expansion is an isenthalpic process. For a rainbow rotating BTZ black hole, the enthalpy is equal to the mass $M$. Therefore, we can obtain the isenthalpic curves in $T-P$ plane by fixing the mass of the black hole. As shown in Fig.4, 5, 6 and 7, the inversion curve divides the $T-P$ plane into two regions. The region above the inversion curve represents the cooling region, with the positive slope. On the other hand, when $J$ and $M$ are fixed, it is found that with the increase of $\eta$, the start point and end point of isenthalpic curve do not change, but its maximum value decreases. This is consistent with the decrease of inversion curve when $\eta$ increases because the inversion curve acts as a boundary between the heating and cooling regions. Thus, the effect of rainbow gravity is to slow down the inverse temperature, meaning that a rainbow rotating BTZ black hole tends to change its heating or cooling action at a lower temperature.

	\section{Discussion and conclusion}\label{444}
	
	In this paper, the Joule–Thomson expansion of a rotating BTZ black hole in rainbow gravity is investigated. Considering the cosmological constant $\Lambda$ as the pressure of the black hole, the Joule–Thomson expansion can describe the expansion of the gas passing through a porous plug or a small valve from high-pressure section to low-pressure section. The enthalpy, which is equal to the mass of the black hole, remains constant during the process. Then, the inversion curves and isenthalpic curves are drawn to determine the heating and cooling regions, which are useful to investigate the effects of rainbow gravity.
	
	\begin{table}[htbp]
		\centering
		\caption{Results and difference between a rainbow rotating BTZ black hole and a rotating BTZ black hole.}
		
		\begin{tabular}{p{4cm}|p{6cm}|p{7cm}}
			\hline
			&Rotating BTZ BH &Rotating BTZ BH in rainbow gravity\\
			\hline
			
			Joule-Thomson coefficient
			&The speed of increase of the Joule–Thomson coefficient is high with a small zero point.
			&The effects of rainbow gravity are to slow down the speed of the increase of the Joule–Thomson coefficient and make its zero point larger. \\
			\hline
			
			Inversion temperature
			&The inversion temperature is high.
			&The effect of rainbow gravity is to slow down the inversion temperature. \\
			\hline
			
			Isenthalpic curve
			&The maximum value of the isenthalpic curve is high.
			&The effect of rainbow gravity is to reduce the maximum value of the isenthalpic curve. \\
			\hline
			
		\end{tabular}
	\end{table}

	We compare the rainbow rotating BTZ black hole and the rotating BTZ black hole without rainbow gravity. The Joule–Thomson expansion of a rotating BTZ black hole has been discussed in \cite{ Liang:2021xny}. To a rotating BTZ black hole without rainbow gravity, $P-V$ critical behavior does not exist. The Joule–Thomson coefficient increases rapidly with a small zero point, and the inversion temperature is high with a large inversion point of isenthalpic curve.
	
	However, to a rainbow rotating BTZ black hole, the rainbow gravity has some influence on Joule–Thomson expansion. The divergence point of the Joule-Thomson coefficient still exists, remains the same value of the black hole without rainbow gravity, and corresponds to the zero point of the Hawking temperature. However, as the value of $\eta$ increases, the effects of rainbow gravity are to slow down the speed of the increase of the Joule–Thomson coefficient and make its zero point larger.
	Meanwhile, the black hole does not have $P-V$ critical behavior, which is similar to rotating BTZ black holes without rainbow gravity. Moreover, the inversion temperature drops with the increase of $\eta$. And when $J$ becomes lager, the effect of rainbow gravity to the inversion temperature becomes smaller. In addition, when $J$ and $M$ are fixed, it is found that with the increase of $\eta$, the start point and end point of isenthalpic curve do not change, but its maximum value decreases. This is consistent with the decrease of inversion curve when $\eta$ increases because the inversion curve acts as a boundary between the heating and cooling regions. Thus, the effect of rainbow gravity is to slow down the inverse temperature, meaning that a rainbow rotating BTZ black hole tends to change its heating or cooling action at a lower temperature, which can be attributed to the topology of black holes.
	
	Herein, we focus on the Joule-Thomson expansion of rotating BTZ black holes. The results show that there are still many other interesting problems worth investigating. Moreover, we compare the existence of critical behavior and the ratio of the minimum inversion in Van der Waals fluid and various black holes, and the results are shown in TABLE I and TABLE II. In future studies, we would further explore the deep relationship between topology and thermodynamics of black holes in rainbow gravity, which might be useful to explore more remarkable effects of the rainbow gravity.

	\section*{Acknowledgement}
	We are grateful to Yuzhou Tao, Haitang Yang and Peng Wang for useful discussions. This work is supported in part by National Science Foundation of China (Grant No. 11747171) and Discipline Talent Promotion Program of Xinglin Scholars of Chengdu University of Traditional Chinese Medicine  (Grant No. QNXZ2018050). The authors contributed equally to this work.

\end{document}